\documentclass[twocolumn,showpacs,preprintnumbers,amsmath,amssymb]{revtex4}

\usepackage{graphicx}
\usepackage{dcolumn}
\usepackage{bm}

\begin{document}


\title{Gamma-rays and neutrinos from dense environment of massive binary systems in open clusters}

\author{W\l odek Bednarek} 
 \email{bednar@uni.lodz.pl}
\author{Jerzy Pabich}%
\author{Tomasz Sobczak}%

\affiliation{%
Department of Astrophysics, The University of \L \'od\'z, 90-236 \L \'od\'z, ul. Pomorska 149/153, Poland
}


\date{\today}

\begin{abstract}
TeV gamma-ray emission has been recently observed from direction of a few open clusters containing massive stars. We consider the high energy processes occurring within massive binary systems and in their dense environment by assuming that nuclei, from the stellar winds of massive stars, are accelerated at the collision region of the stellar winds. We calculate the rates of injection of protons and neutrons from fragmentation of these nuclei in collisions with stellar radiation and matter of the winds from the massive companions in binary system. Protons and neutrons can interact with the matter, within the stellar wind cavity and within the open cluster, producing pions which decay into $\gamma$-rays and neutrinos. We discuss the detectability of such $\gamma$-ray emission by the present and future Cherenkov telescopes for the  case of two binary systems Eta Carinae, within the Carina Nebula, and WR 20a, within the Westerlund 2 open cluster. We also calculate the neutrino fluxes produced by protons around the binary systems and within the open clusters. This neutrino emission is confronted with ANTARES upper limits on the neutrino fluxes from discrete sources and with the sensitivity of IceCube.
\end{abstract}

\maketitle

\section{Introduction}

Open clusters have been suspected as sites for the high energy processes due to the presence of large concentration of matter and also early type compact objects such as massive stars, compact massive binaries, pulsars and their nebulae and supernova remnants. 
In fact, up to now TeV $\gamma$-ray emission has been detected from the direction of 3 open clusters (i.e. Cyg OB 2 - Aharonian et al.~2002, Westerlund 2 - Aharonian et al.~2007, and Westerlund 1 - Abramowski et al.~2012a). Also GeV $\gamma$-ray emission, extending up to $\sim$100 GeV, has been recently reported from the direction of the supermassive binary system Eta Carinae within the Carina Nebula complex (Tavani et al.~2009a, Abdo et al. 2009a, Farnier et al.~2011). A part of this GeV emission shows evidence of modulation with the binary period (Farnier \& Walter~2011, Reitberger et al.~2012). Therefore, it has to be produced not far from the binary system itself. Eta Carinae binary system has not been detected up to now in the TeV $\gamma$-ray range (Abramowski et al.~2012b). 
The modulated GeV $\gamma$-ray emission from binary system Cyg X-3 has been recently detected by the AGILE and Fermi-LAT telescopes (Tavani et al.~2009b, Abdo et al.~2010, Bulgarelli et al.~2012).
This emission is transient and appears close to the quenched radio state just before the major radio flares observed from this system.  
The GeV $\gamma$-ray emission has not been identified up to now in the Fermi-LAT data from the binary system WR 20a within the open cluster Westerlund 2. Two new sources reported by Fermi collaboration in this direction have been identified with the $\gamma$-ray pulsars (Abdo et al. 2009b, Saz Parkinson et al.~2010).

A few scenarios have been considered as possible explanations of the high energy emission from the open clusters.
It has been proposed that strong winds produced by massive stars in open clusters are able to accelerate particles at shock waves
formed during collisions of the winds with dense matter of the open cluster (e.g. V\"olk \& Forman~1982, Cesarsky \& Montmerle~1983, Giovannelli et al.~1996, Torres et al.~2004, Bednarek 2007). $\gamma$-rays and neutrinos can be also produced by particles accelerated at the shocks formed within massive binary systems (e.g. Eichler \& Usov~1993, White \& Chen~1995, Benaglia \& Romero 2003, Bednarek~2005a, Reimer et al.~2006, Pittard \& Dougherty~2006, Bednarek \& Pabich~2011). Massive stars within open clusters end their life at a relatively short time producing supernova remnants (SNRs) and pulsar wind nebulae (PWNe). Interaction of SNRs shock waves with dense clouds within the open cluster can also accelerate relativistic particles that are able to produce high energy radiation (Aharonian \& Atoyan~1996).
A part of $\gamma$-ray emission from the open clusters can be also produced within PWNe or as a result of the interaction of particles accelerated within PWNe with dense matter of an open cluster (Bednarek~2003, Bartko \& Bednarek~2008, Abramowski et al.~2011, Ohm et al.~2013, Aliu et al.~2014).  It is not clear at present which of those scenarios provide a bulk of the high energy $\gamma$-ray emission observed from a few open clusters mentioned above. It is likely that in specific open cluster a few processes can be important since different types of objects (massive star winds, SNRs, PWNe) might inject relativistic particles with comparable power, i.e. of the order of $10^{50}$ erg. 

In the present paper we investigate in detail the high energy radiation expected in hadronic processes within and around the massive binary systems surrounded by the large concentration of matter. It is assumed that nuclei are accelerated in the region of colliding winds within the binary stars (e.g. see Bednarek~2005b). These nuclei can severely  disintegrate, in the interaction with the radiation field of massive stars and with the matter of the stellar winds, injecting neutrons and protons. Charged protons diffuse through the open cluster producing $\gamma$-rays and neutrinos in collisions with the matter of the stellar wind cavity and dense environment of the open cluster. On the other hand, neutrons move balistically through the wind cavity and decay into protons at some distance from the binary system which can be still within the wind cavity or already within the open cluster. Protons from their decay can also contribute to the high energy $\gamma$-ray and neutrino spectrum.
As an example, we perform calculations of the $\gamma$-ray and neutrino fluxes produced in the clusters surrounding the Eta Carina supermassive binary system and WR 20a binary system in Westerlund 2 open cluster.
This emission is confronted with the present measurements of the GeV-TeV $\gamma$-ray emission from these clusters, sensitivities of the future Cherenkov Telescope Array (CTA) and the IceCube neutrino telescope.

\section{A massive binary system within the open cluster}

We consider a massive binary system in which one or both companion stars belongs to the class of the Wolf-Rayet (WR) stars.
WR type star produces fast and dense wind, due to the huge mass loss rate, which can be of the order of ${\dot M_{\rm WR}} = 10^{-5}{\dot M}_{\-5}$ M$_\odot$ yr$^{-1}$. The winds propagate with the characteristic velocities of the order of $v_{\rm w} = 10^3v_3$ km s$^{-1}$.
The density of the wind drops with the distance from the star according to, 
\begin{eqnarray}
n_{\rm w}(r)\approx 3.2\times 10^{11}{\dot M}_{-5}/v_3R_{12}^2r^2
~~~{\rm cm^{-3}},   
\label{eq1}
\end{eqnarray}
\noindent
where $R_{\rm WR} = 10^{12}R_{12}$ cm is the radius of the star, and $r=R/R_{\rm WR}$ is the distance from the star in units of the stellar radius. Note that at the distance of the order of a parsec, density of the wind becomes very low, e.g. for 1 pc the wind density becomes $\sim 0.03$ cm$^{-3}$. So then, the outer regions of the wind cavity are filled with very rare but high velocity gas.

The massive binary systems are usually immersed itself within a relatively dense open clusters (OCs). Typical densities of the OCs are of the order of $n_{\rm oc} = 10n_{10}$ cm$^{-3}$ and temperatures of the gas of the order of $T_{\rm oc} =10^4T_4$ K. At certain distance from the binary system, the pressure of the stellar wind is balanced by the pressure of thermal gas within the OC. We estimate the dimension of such stellar wind cavity, in the case of a constant wind velocity, by comparing the energy density of the wind with the energy density of the medium within the OC. This allows us to estimate the radius of the wind cavity, 
\begin{eqnarray}
R_{\rm cav}\approx 1.1\times 10^{19}[{\dot M}_{-5}v_3/(n_{10}T_4)]^{1/2}~~~{\rm cm}.   
\label{eq2}
\end{eqnarray}
\noindent
Note, that the stellar wind at first accelerate from the stellar surface  reaching the asymptotic velocity (see Waters et al.~1988). 
After initial acceleration, due to the gas pressure, the wind decelerates due to the gravitational attraction of the massive star. The velocity of the wind drops with a distance from the star according to,
$v^2_{\rm w}(R) = v^2_{\rm o} - 2GM_{\rm wr}(1/R_{\rm WR} - 1/R)\approx
10^{16}v_3^2 - 2.7\times 10^{15}(M_{10}/R_{12})(1 - 1/r)$ cm$^2/s^2$,
where $M_{\rm WR} = 10M_\odot M_{\rm 10}$ is the mass of the star and $G$ is the gravitational constant. For the parameters of the stars considered in this paper, the deceleration effect of the wind can be neglected.
The radii of the wind cavities around WR type binary systems within the open clusters are typically of the order of a few parsecs for the density of surrounding matter of the order of $\sim 10$ cm$^{-3}$ and its temperature $T_{\rm oc}\sim 10^4$K.
In fact, the scenario discussed above is a simplification of the real situation in which the interaction of the stellar winds with
open cluster matter can result in a complicated double shock structure. Additionally the cavity does not have to be isotropic due to anisotropies of the environment.
We expect that the above estimated radius of the wind cavity is approximately correct since the amount of the mass overtaken by the expanding stellar wind (with the radius of the order of a few pc) is much smaller than the total mass of the open cluster (with the radius of several pc). Therefore, the environment in which the binary system is immersed does not change significantly with time.

We consider consequences of acceleration of nuclei within the binary system located in the open cluster. Nuclei (from the stellar winds) can be accelerated within the collision region of the winds (e.g. Eichler \& Usov~1992). The recconnection of the magnetic field and the diffusive shock acceleration process can play important role in this place. The details of the acceleration process, propagation of hadrons and their subsequent interaction within the binary system and its surrounding are discussed in the next section. We show that nuclei can efficiently disintegrate in the dense radiation and matter of the winds from massive stars. As a result, neutrons are injected. They decay at the distance from the binary system which is determined by their Lorentz factors.
Protons, extracted from nuclei, lose energy on interaction with the dense wind close to the binary system. They also suffer adiabatic energy losses in the expanding wind within the wind cavity. 
We consider high energy processes in which $\gamma$-rays and neutrinos are produced in hadronic collisions in the above described scenario. The schematic representation of the processes around the binary system is shown in Fig.~1.

Note that hadrons might be also accelerated in such scenario at the shock structure formed in the collision of the stellar wind
with the circumstellar gas. We expect that this acceleration process is less efficient than the acceleration process occurring within the binary system. The product of the magnetic field strength and the characteristic dimension of the shock, determining the maximum energies of accelerated particles, is expected to be lower at the shock produced by the wind in surrounding gas due to the mainly radial of the magnetic field within the binary system and its toroidal structure far away from the binary.

\begin{figure}
\vskip 7.5truecm
\includegraphics{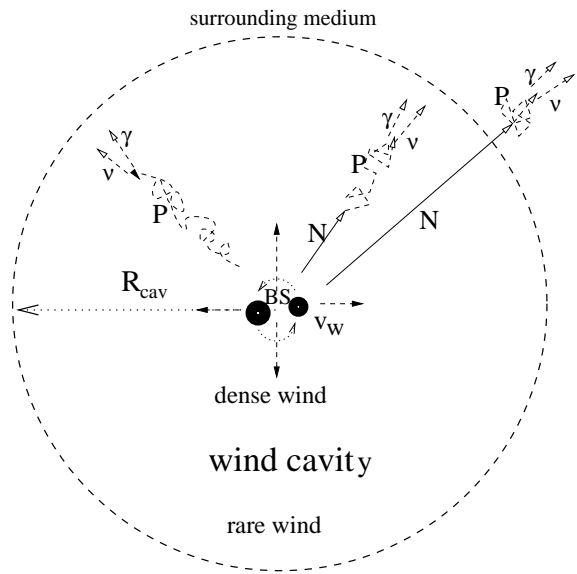}
\caption{Schematic representation of the massive binary system. Powerful stellar wind creates a cavity in the medium of the open cluster. A cavity has the radius $R_{\rm cav}$. It is filled with dense wind, close to the binary system, and a rare wind at large distances. The wind is expanding
with the velocity $v_{\rm w}$. Relativistic nuclei are accelerated in the region of colliding winds of the massive stars. These nuclei disintegrate in collisions with the stellar radiation and the matter of the wind, injecting relativistic protons and neutrons. Protons (P) are captured in the expanding wind suffering huge adiabatic energy losses. They can also lose a part of their energy on collisions with the matter of the wind. High energy $\gamma$-rays and neutrinos are produced in these hadronic collisions.
Neutrons (N), from fragmentation of nuclei, propagate along the straight lines and decay at some distance from the binary system. Protons, from decaying neutrons, are captured by the local magnetic field. They can interact efficiently with the surrounding matter
producing $\gamma$-rays and neutrinos. Since the adiabatic energy losses of these protons, from decaying neutrons, are relatively low, produced by them $\gamma$-rays and neutrinos have larger energies than those produced by protons from direct fragmentation of primary nuclei.}
\label{fig1}
\end{figure}
\section{Injection of nuclei from the binary system}

Massive stars produce fast and dense winds which can extract significant amount of original mass from the star during its lifetime. In fact, the intensive mass loss rates of the order of a few $10^{-6}-10^{-4}$ M$_\odot$ yr$^{-1}$ are characteristic for the Wolf-Rayet (WR) and O type stars. During the main sequence stage, the outer parts of stars are completely lost and only inner parts, composed of heavy nuclei, are left. Therefore, the winds of early type stars are expected to be mainly composed from nuclei heavier than hydrogen such as helium to oxygen. 

Massive stars are frequently found within the massive binary systems in which strong winds collide providing conditions for acceleration of nuclei to large energies.
Below, we generally consider the process of acceleration of nuclei at the shocks or the reconnection regions created by these colliding winds. The propagation and interaction of nuclei with the stellar radiation field results in their photo-disintegration to neutrons, protons and secondary nuclei.
We perform numerical simulations of nuclei in the radiation field of the massive stars in order to determine the rate of injection of different nuclei. Note that density of stellar photons in the vicinity of massive star, $n_{\rm ph}\approx 2\times 10^{16}T_5$ ph. cm$^{-3}$ (where $T = 10^5T_5$ K is the surface temperature of the massive star), is a few orders of magnitude larger than density of matter in the stellar wind (see Eq.~\ref{eq1}). Therefore, it is expected that nuclei with sufficiently large energies interact at first in collisions with stellar photons rather than with the matter of the stellar wind. After that, nuclei can suffer significant fragmentation in collisions with the matter of the stellar wind.
This second process is independent on energy of nuclei. Therefore,
lower energy nuclei can also suffer strong disintegration process in collisions with dense matter of the stellar winds in the case of stars with exceptionally strong  winds such as considered in this paper (i.e. WR and O type stars which mass loss rate is above $\sim 10^{-5}$ M$_\odot$ yr$^{-1}$). Neutrons released from nuclei in collisions with the matter have the spectrum similar to the spectrum of primary nuclei. Therefore, neutrons with low energies can decay relatively close to the binary system where the density of matter is still high. On the other hand, high energy neutrons can even reach dense regions outside the wind cavity of the massive binary system. In this paper we calculate $\gamma$-ray and neutrino spectra produced by protons extracted from nuclei and also from protons from neutrons decaying at some distance from the binary system. These neutrons decay within the wind cavity and also within the dense open cluster in which binary system is immersed.

We estimate the maximum energies of hadrons accelerated in the shock region of colliding winds following the conditions considered e.g. in Bednarek \& Pabich~(2011) and Bednarek~(2005b). The maximum energies of hadrons are determined by comparing their acceleration time scale and their 
escape time scale from the acceleration region or collision time scale with the matter of the stellar wind.  
The acceleration time scale is given by, 
\begin{eqnarray}
\tau_{\rm acc} = E_{\rm h}/\dot{P}_{\rm acc}\approx 0.02\gamma_{\rm h}/(\chi_{-5}B_3)~~~{\rm s}, 
\label{eq3}
\end{eqnarray}
\noindent
where $E_{\rm h}$ and $\gamma_{\rm h}$ are the energy and the Lorentz factor of particles, $\dot{P}_{\rm acc} = \chi cE_{\rm h}/R_{\rm L}$ is the acceleration rate,
$\chi = 10^{-5}\chi_{-5}$ is the acceleration coefficient, $R_{\rm L} = Ac\gamma_{\rm h}/ZeB_{\rm sh}$ is the Larmor radius of hadron in the magnetic field at the shock $B_{\rm sh}$, $c$ is the velocity of light, $e$ is the elemental charge, and $A$ and $Z$ are the mass and charge numbers of nuclei. We apply $A/Z = 2$. $B_{\rm sh}$, is related to the surface magnetic field of the massive star by assuming its dipolar structure close to the surface up to $\sim 1.2R_\star$ and radial structure at larger distances (Usov \& Melrose~1992)). For the distance of the collision region from the star equal to $R_{\rm sh} = 2 R_\star$, $B_{\rm sh}$ drops to $\sim 0.25B_\star$. 

The advection time scale of hadrons from the collision region is estimated from, 
\begin{eqnarray}
\tau_{\rm adv}\approx R_{\rm sh}/v_{\rm w}\approx 10^4R_{12}/v_3~~~{\rm s}, 
\label{eq4}
\end{eqnarray}
\noindent
where $R_{\rm sh} =10^{12}R_{12}r_{\rm sh}$ cm is the distance of the collision region from the star and $r_{\rm sh} = R_{\rm sh}/R_{\rm WR}$. We estimate the diffusion time scale of hadrons (in the Bohm approximation) in the acceleration region  $\tau_{\rm dif}\approx 1.5\times 10^4R_{12}^2B_3/\gamma_6$ s, where $\gamma = 10^6\gamma_6$ is the Lorentz factor of hadrons and $B_{\rm sh} = 10^3B_3$ G. This diffusion time is larger than the above estimated advection time scale even for the most energetic hadrons considered in this scenario.

Then, the maximum energy of hadrons accelerated at the region of colliding stellar winds is obtained from comparison of Eq.~3 and Eq.~4. It is then given by,
\begin{eqnarray}
\gamma_{\rm max}^{\rm adv}\approx 5\times 10^5 B_3v_3R_{12}r_{\rm sh}~~~{\rm GeV},   
\label{eq5}
\end{eqnarray}
\noindent
where the acceleration coefficient is estimated to be $\chi = (v_{\rm w}/c)^2\approx 10^{-5}v_3^2$.

\begin{table*}[t]
  \caption{Parameters of massive stars in binary systems,  maximum Lorentz factors of accelerated nuclei and the optical depths for protons in the matter of the stellar wind}
  \begin{tabular}{llllllllllll}
\hline 
\hline 
\\
Name   & B &  R$_{\rm WR}$  &  $v_{\rm w}$ & R$_{\rm sh}$  & T$_{\rm WR}$ & $M_{\rm WR}$ & ${\dot M}_{\rm WR}$ &  $L_{\rm w}$ &  $\tau_{\rm hp}$ & $\gamma_{\rm max}^{\rm adv}$ & $\gamma_{\rm max}^{\rm col}$ \\
unit    &   G   &  cm & km/s     &  $R_{\rm WR}$  & K & M$_\odot$ &
 M$_\odot$/yr  &  erg s$^{-1}$ &  &  \\
\hline
\\
WR   &  $3\times 10^3$~~~  & $10^{11}$           & $3\times 10^3$ & 2  &  $1.4\times 10^5$~~  & 5  & $3\times 10^{-6}$  &   $10^{37}$    & $\sim$0.3  &  $9\times 10^5$ &  $3.8\times 10^6$\\
\\
WR 20a   &  $10^3$          & $1.4\times 10^{12}$ & $10^3$         & 2  &  $4\times 10^4$ & 30 & $3\times 10^{-5}$  &  $10^{37}$ & $\sim$2.4  & $1.4\times 10^6$ &
$10^6$ \\
\\
Eta Car  &  $200$          & $1.2\times 10^{13}$           & $700$          & 1.4  &  $4\times 10^4$ & 80 & $2.5\times 10^{-4}$~~  &  $4\times 10^{37}$  & $\sim$4.1~~  & $1.2\times 10^6$~~  &  $2.7\times 10^5$ \\
\hline
\hline 

\end{tabular}
  \label{tab1}
\end{table*}

On the other hand, hadrons lose also energy on collisions with the matter of the stellar wind. The hadron-hadron energy loss time scale can be estimated from,
\begin{eqnarray}
\tau_{\rm hh} = (cn_{\rm w}(r)k\sigma_{\rm pp})^{-1}\approx 3.5\times 10^3 R_{12}^2r^2v_3/{\dot M}_{-5}~~~{\rm s},
\label{eq6}
\end{eqnarray}
\noindent
where $\sigma_{\rm pp} = 3\times 10^{-26}$ cm$^2$ is the cross section for proton-proton collision, and $k = 0.5$ is the in-elasticity coefficient.

The maximum energies of hadrons allowed by collisional energy losses are,
\begin{eqnarray}
\gamma_{\rm max}^{\rm hh}\approx 3.5\times 10^5R_{12}^2r^2v_3^3B_3/{\dot M}_{-5}.
\label{eq7}
\end{eqnarray}
\noindent
We have calculated the maximum energies of hadrons due to these two processes for three example massive stars within the binary systems. The results are shown in Table~1. In the case of the parameters of the WR type star as observed in Cyg X-3 binary system, the maximum energies of accelerated hadrons are limited by advection process. In the case of the binary system containing WR 20a stars, the maximum energies of hadrons are comparable in the case of advection process and collisional energy losses. But, in the case of Eta Carinae the maximum  energies of hadrons are limited by collisional energy losses. In all considered cases hadrons can reach energies of the order of $\sim$10$^6$ GeV.

In all calculations of the $\gamma$-ray and neutrino spectra (presented below), it is assumed that nuclei are injected with the power law spectrum (spectral index equal to 2), extending up to the lower value between those given by Eqs.~5 and 7. The power in these hadrons is normalized to a part, $\eta$, of the power provided by the stellar wind of the massive star, $L_{\rm h} = \eta L_{\rm w}$ (see Table~1 for specific binaries). In Table~1 we also report other basic parameters of these three binary systems. Their orbital periods are equal to 4.8 hrs for Cyg X-3 binary system  (Becklin et al. 1973), 3.675 days for WR 20a (Rauw et al.~2004) and 2027.7 days for Eta Carinae (Damineli et al.~2008)).

\section{Disintegration of hadrons within binary system}

The process of disintegration of nuclei has been extensively studied in the past in the context of the origin and propagation of the highest energy cosmic rays (e.g. Stecker~1969, Tkaczyk et al.~1975, Puget et al.~1976).
This process has been also proposed to be important in compact sources in which dense soft radiation field is expected, i.e. in supernova remnants (e.g. Pollack \& Shen~1969), binary systems (e.g. Karaku\l a et al. 1994), dense stellar clusters (e.g. Anchordoqui et al.~2007a,b), or the Galactic Center (e.g. Kusenko et al.~2006).
$\gamma$-ray emission from de-excitation of fragments of nuclei, which survived disintegration, has been calculated in some of these works following the prescription emphasized by Balashov et al.~(1990).
This $\gamma$-ray production process is generally relatively inefficient since only a small part of the initial energy of hadrons is converted into radiation. However, it may play a role in the case of sources with strong
soft radiation field but without large amount of matter.
Here we consider production of $\gamma$-rays and neutrinos by secondary products of disintegrated nuclei
in their interaction with dense matter surrounding the source of relativistic nuclei. We neglect possible direct production of neutral high energy radiation in the interaction of hadrons with the stellar radiation ($p + \gamma\rightarrow \pi$) since, as we have shown above, nuclei are not accelerated to large enough energies to reach the threshold for this process in the environment of the binary systems.

In order to have an impression about the role of the fragmentation process of nuclei in the radiation field of massive stars we consider a simple case of their injection from a  point-like source not far from the stellar surface, e.g. at the distance of two stellar radii. We calculate the optical depths for fragmentation of nuclei injected at different initial directions, as a function of their energy and two mass numbers corresponding to the helium and the oxygen nuclei (see Fig.~2). The approximation of the cross section for photo-disintegration process of nuclei is applied as described in Karaku\l a \& Tkaczyk~(1993).
Nuclei are considered to propagate in the vicinity of three representative massive stars with different parameters: WR type star as observed in the Cyg X-3 binary system, WR star in the WR 20a binary system and supermassive star in the Eta Carinae binary system (see Fig.~2). Note that only nuclei with Lorentz factors above $\gamma_{\rm min}$
can be efficiently disintegrated. This critical Lorentz factor can be estimated from $\gamma_{\rm min} = E_\gamma^{\rm th}/3k_{\rm B}T\sim 4\times 10^4/T_5$, where $E_\gamma^{\rm th} = 2$ MeV is the minimum energy of stellar photon in the nuclei rest frame which is able to effectively photo-disintegrate the nuclei, $T = 10^5T_5$ K is the surface temperature of the star, and $k_{\rm B}$ is the Boltzmann constant. In this figure we also determine the number of nucleons dissolved from the primary nuclei in the photo-disintegration process. The investigation of the results presented in Fig.~2, in the context of the maximum energies of nuclei expected in our model (see Table.~1), allows us to conclude that in the case of WR 20a binary system the process of photo-disintegration of nuclei can be neglected. In the case of WR star with parameters observed in Cyg X-3 binary system and Eta Carinae, significant number of nuclei with energies in the range $\sim 10^{(5-6)}$ will lose nucleons in collisions with stellar radiation.

In order to conclude whether nuclei can be efficiently disintegrated in collisions with the matter of the stellar wind, we estimate the optical depth for relativistic hadrons on collisions with the matter of the wind. This optical depth can be estimated from,
\begin{eqnarray}
\tau_{\rm hp}=\int_{R_{\rm BS}}^{R_{\rm c}}{{\sigma_{\rm pp}n_{\rm w}c}\over{v_{\rm w}}}dR\approx 
2.9\times 10^{12}{{{\dot M}_{-5}}\over{v_3^2}}({{1}\over{R_{\rm BS}}}-
{{1}\over{R_{\rm c}}}),
\label{eq8}
\end{eqnarray}
\noindent
where $R_{\rm BS}$ (in cm) is the radius of the binary system at which hadrons are injected within the wind. It is assumed to be equal to the shock radius.
The optical depths for protons on the interaction with the matter of the wind are shown in Table~1 for three example parameters of the massive stars. Note that the optical depth for nuclei with specific mass number $A_1$ in collisions with nuclei with the mass number $A_2$ scales as, $\tau_{A_1A_2}\propto (A_1A_2)^{\beta}$, where $\beta$ is between $2/3$ to $1$ (see DPMJET calculations given in Mori~2009). For the case of the helium nuclei ($A_1 = A_2 = 4$), the optical depths will be a factor of $\sim 14$ larger ($\beta = 0.94$) and for the oxygen nuclei a factor of $\sim 43$ larger ($\beta = 0.68$, see Mori~2009). Therefore, we conclude that in the case of considered binary system relativistic nuclei should be completely disintegrated.
Large values of the optical depths for collisions of protons with the matter of the stellar winds indicate that these relativistic hadrons should efficiently lose energy on production of high energy $\gamma$-rays and neutrinos in the close vicinity of the massive binary system. On the other hand, neutrons, released from these nuclei, should move rectilinearly and decay at large
distances from the binary system within the wind cavity or, in the case of the most energetic neutrons, directly in the region of the open cluster. Protons, from decay of these neutrons, can also produce high energy radiation at large distances from the binary systems, i.e in the region with low level of the soft radiation field and low density of matter in respect to the density of the wind within the binary system.

\begin{figure}
\vskip 8.7truecm
\includegraphics{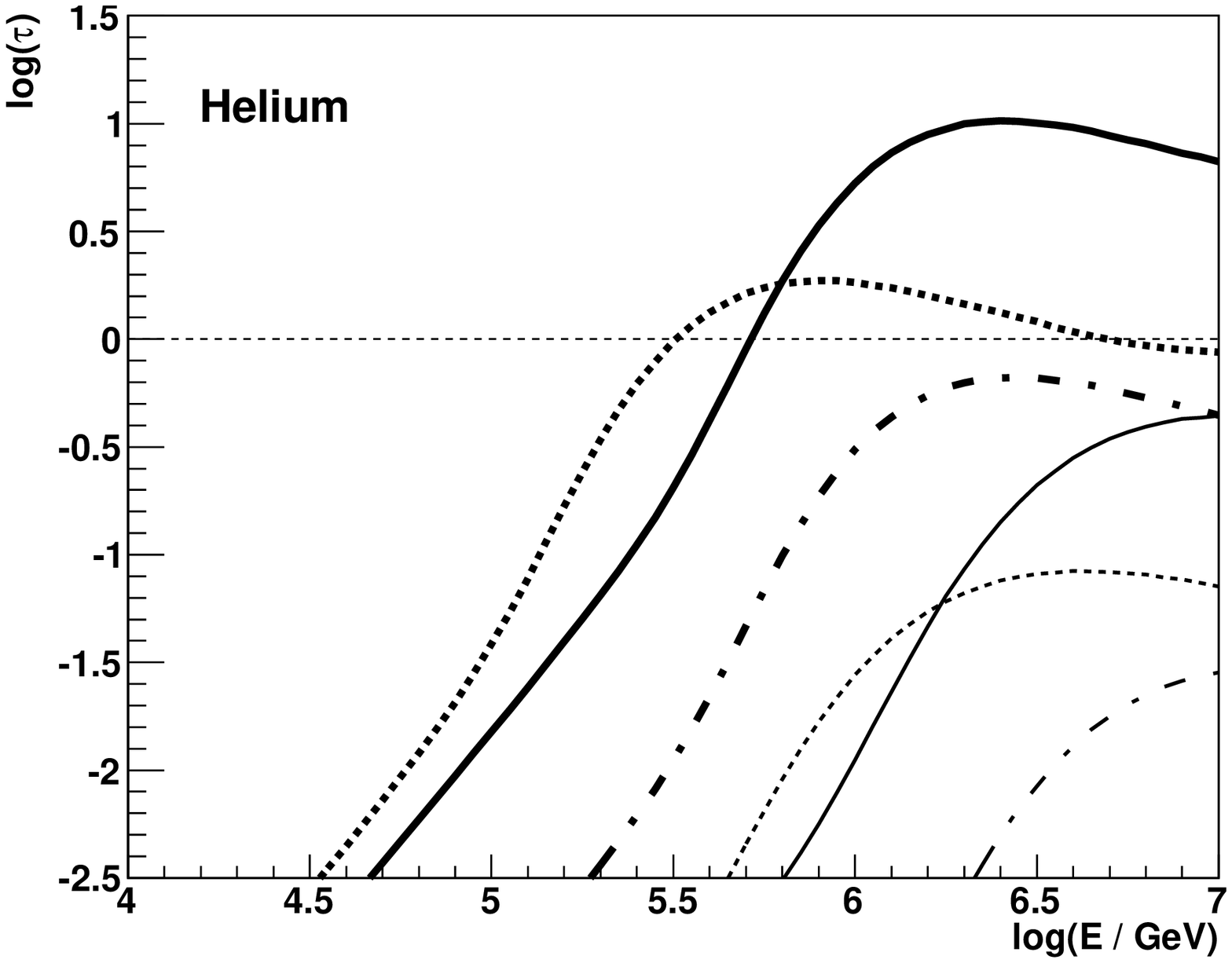}
\includegraphics{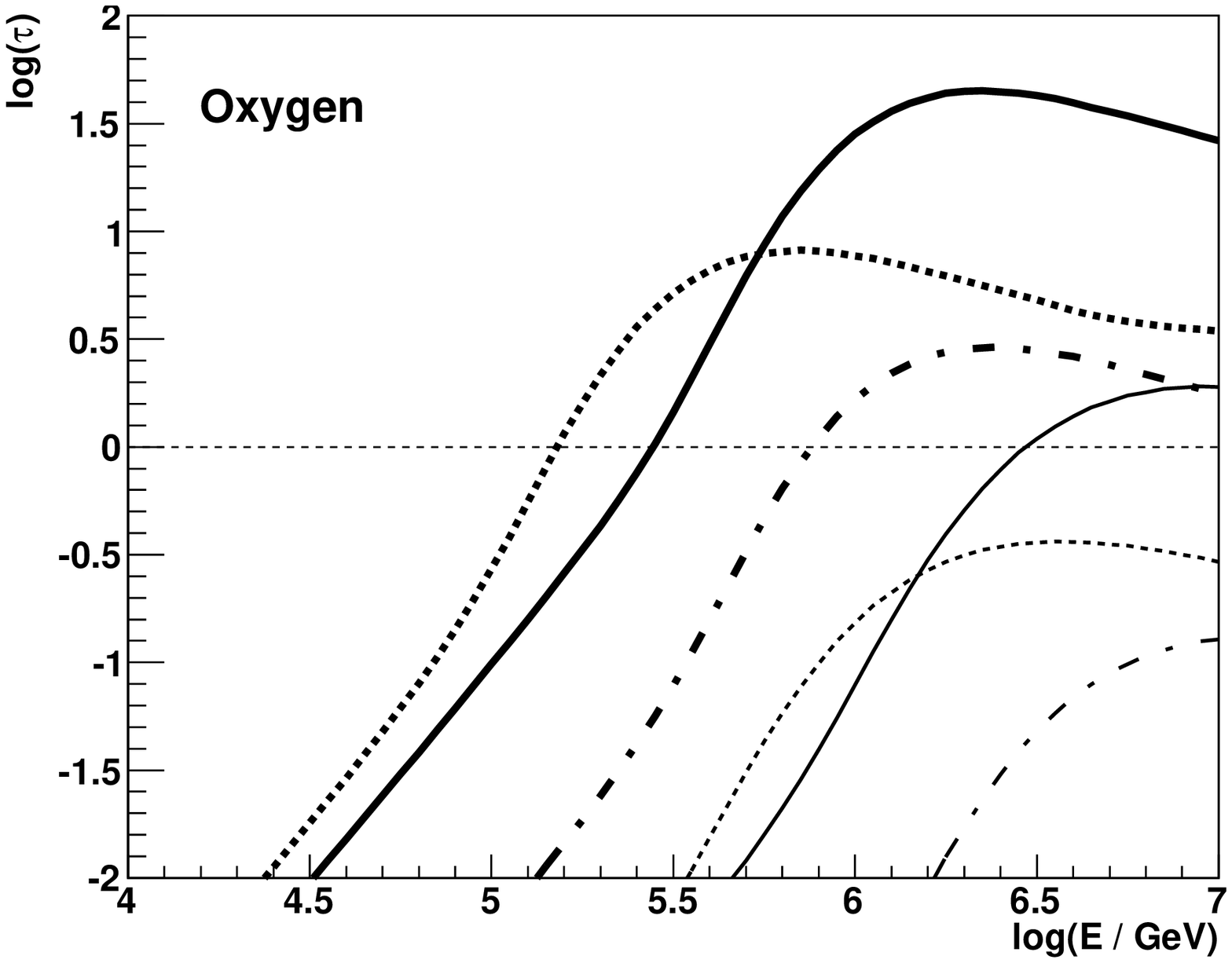}
\includegraphics{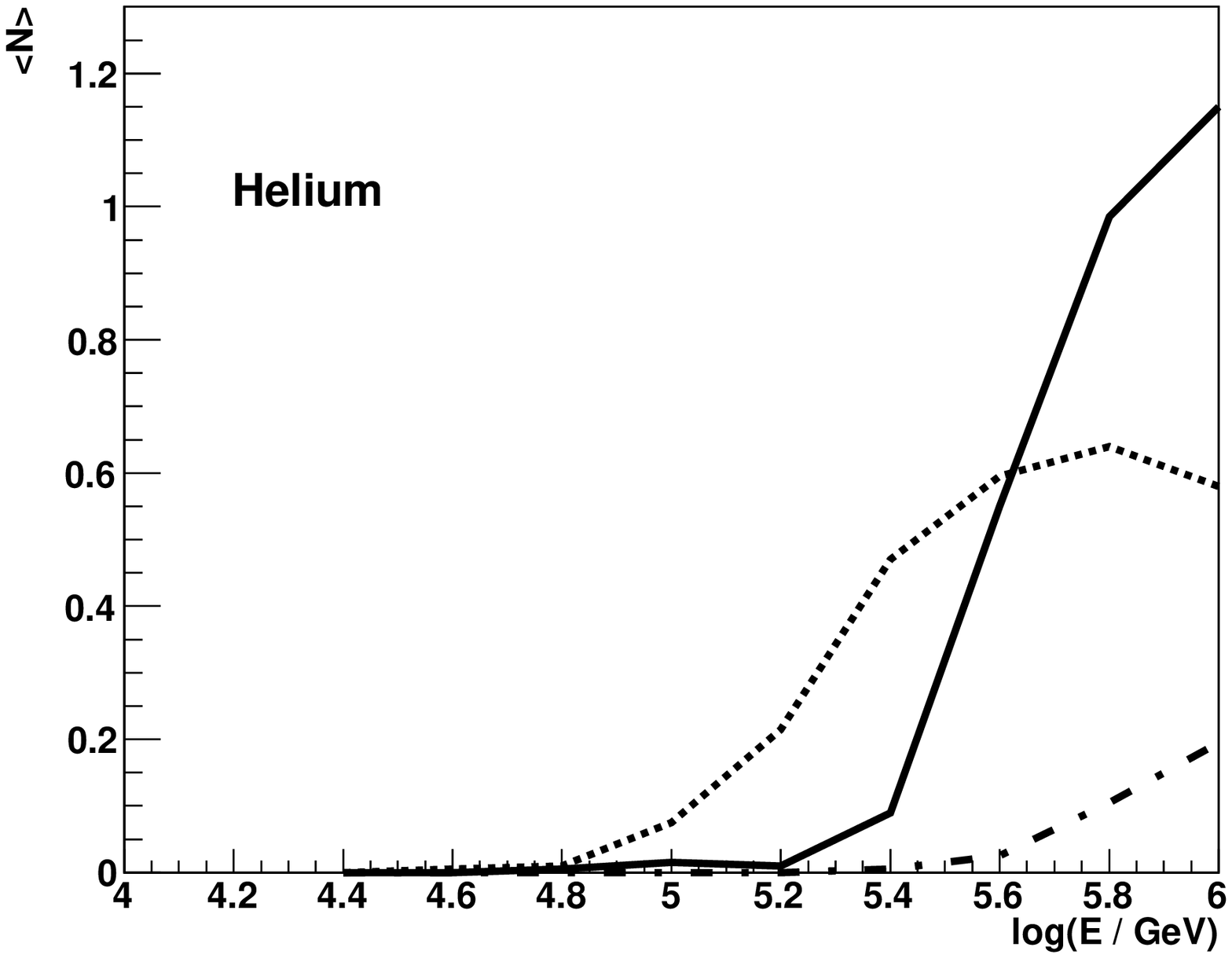}
\includegraphics{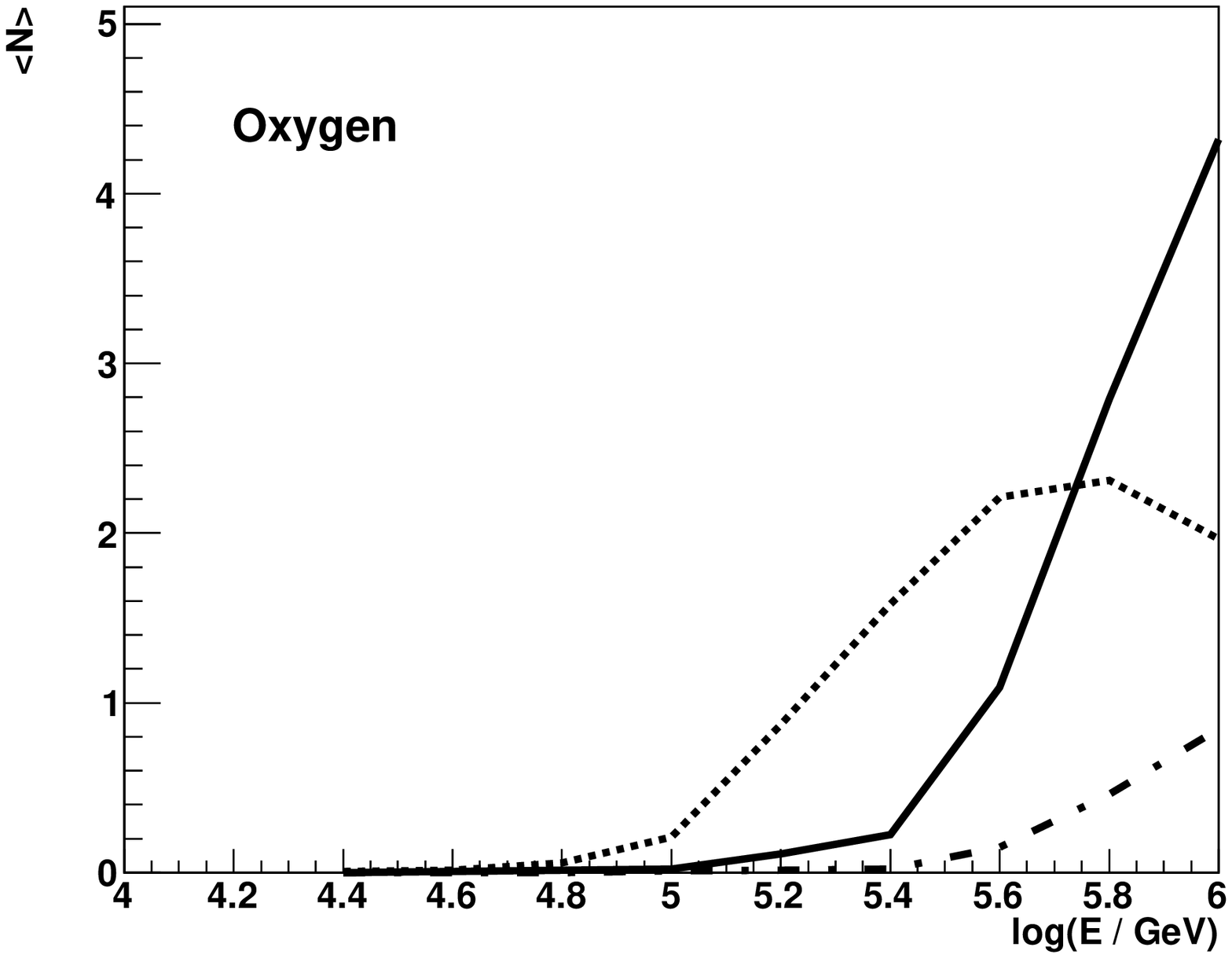}
\caption{Upper panel: The optical depths for nuclei, as a function of their energy (per nucleon), for two injection angles $\alpha = 45^o$ (thick curves) and $135^o$ (thin curves)
(measured from the direction towards the star) for the helium and oxygen nuclei and for three considered massive stars: WR in Cyg X-3 binary system (dotted), WR 20a binary system (dot-dashed) and Eta Carinae (solid). 
Lower panel: The average number of neutrons extracted from relativistic nuclei (as a function of their energy per nucleon) with different initial mass numbers, He (left figures), O (right) as a result of their photo-disintegration in the stellar radiation field for three considered example massive stars: WR type star as observed in Cyg X-3 (dotted curve), WR 20a (dot-dashed) and Eta Carinae (solid). Nuclei are injected isotropically at the distance of the shock from the star (see Table~1).}
\label{fig2}
\end{figure}
\section{Hadrons escaping from the binary system}

Accelerated nuclei initiate a sequence of processes within the binary system and its surrounding.
As we have shown above nuclei suffer complete disintegration if injected within massive binary systems containing luminous stars
characterized by dense stellar winds (WR type stars). Unstable neutrons, from photo-disintegration of these nuclei, decay at some distance from the binary system.
Depending on their energy, they can decay within the stellar wind region (the wind cavity) or outside the stellar wind shock, i.e.
within the open cluster. Protons, from neutrons decaying within the stellar wind, are expected to suffer adiabatic energy losses during the fast expansion of the wind. These protons are also partially advected with the wind to the open cluster. Neutrons with large enough energies can also decay outside the wind cavity in the volume of the open cluster. Protons from their decay diffuse gradually through the open cluster suffering some collisions with the distributed matter.
On the other hand, primary nuclei, accelerated within the binary system, and also protons from their fragmentation are captured by the magnetic field in the dense stellar wind. These hadrons can interact with the matter of the wind. As a result of these interactions, an additional population of neutrons is produced not far from the vicinity of the binary system where the matter in the stellar wind is still relatively dense. These neutrons decay within or outside the wind cavity as described above.

We conclude that the simple scenario which postulates acceleration of hadrons within the massive binary system immersed in the open cluster provides variety of conditions for production of high energy radiation in different parts of open cluster characterized by different conditions, e.g the wind cavity of the binary system and dense medium of the open cluster. We are interested in the $\gamma$-ray and neutrino emission produced in such scenario 
in the context of recent observations of the TeV $\gamma$-ray emission from open clusters (e.g. Westerlund 2 and the Carina Complex). 

Let us at first estimate whether charged hadrons (accelerated within the binary system and products of their fragmentation) can be captured in the stellar wind. We compare the Larmor radius of charged hadrons
with a specific energy with the characteristic distance scale which is the distance at which hadrons are located from the binary system, $R$. The Larmor radius of hadrons with the Lorentz factor $\gamma_{\rm n}$
is given through the expression $R_{\rm L} = 3\times 10^6\gamma_{n}/B_{\rm G}$ cm, where $B = 1B_{\rm G}$ G is the magnetic field strength in the wind at the distance $R$ from the binary system. The magnetic field in the stellar wind is expected to have complicated structure as a function of the distance from the star (dipolar, radial and at farther distances toroidal). This structure becomes even more complicated in the case of stars within the compact binary system when both stars have large velocities. We assume that the magnetic field is radial at distances below $\sim 10R_{\rm WR}$, i.e. in this region $B\propto B_\star (10R_{\rm WR}/R)^2$. At larger distances the magnetic field becomes toroidal.
Then, for distances greater than $\sim 10R_{\rm WR}$, the magnetic field strength can be approximated by
$B(R)\approx 100B_3(R_{\rm WR}/R)$ G, where $B = 10^3B_3$ G is the surface magnetic field of the WR type star. Applying this simple scaling, the condition for capturing of protons, $R_{\rm L}<R$, is fulfilled for $\gamma_{\rm n} < 3\times 10^6$. Protons with energies fulfilling this condition are captured in the stellar wind. They are expected to suffer strong adiabatic energy losses during gradual expansion of the wind from the star.
On the other hand, hadrons with larger energies can leave the stellar wind region (and the wind cavity) without significant energy losses. They also propagate almost along the straight lines without significant collisions with the matter of the wind.
However, hadrons with such large energies are not expected to be accelerated at the collision region within the binary system (see Table 1).

Hadrons captured in the stellar wind lose energy on the adiabatic process. We can determine the Lorentz factor of hadrons at a specific distance from the binary system by taking into account adiabatic and collisional energy losses,
\begin{eqnarray}
\gamma_{\rm h}(R) = \gamma_{\rm h}(R_{\rm BS})
R_{\rm BS}k^{\tau_{\rm hp}}/R,
\label{eq9}
\end{eqnarray}
\noindent
where $\tau_{\rm hp}$ is given by Eq.~\ref{eq8}, and $k\approx 0.5$ is the in-elasticity coefficient in proton-proton collisions. 
Depending on the parameters of the considered scenario, either adiabatic losses or collision losses determine the Lorentz factors of hadrons at a specific distance of the wind from the binary system.
The Lorentz factors of hadrons at a specific distance from the binary system, in the case of only collisional and also collisional and adiabatic energy losses, for three example stars within the binary systems $\eta$ Car, WR 20a and
WR star in Cyg X-3, are shown in Fig.~3. There are huge differences in the values of these Lorentz factors with and without adiabatic energy losses. Therefore, adiabatic energy losses play an important role in the process of production of high energy radiation in the open clusters.

\begin{figure}
\vskip 6.2truecm
\includegraphics{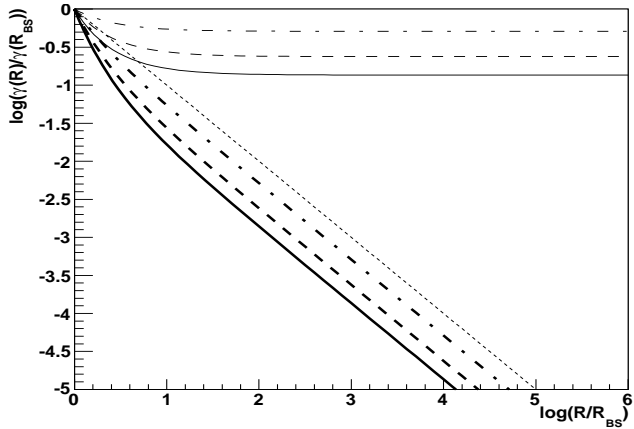}
\caption{The change of the Lorentz factor of a charged hadron as a function of distance from the binary system caused by only collisional energy losses (thin curves) and by both collisional and adiabatic energy losses (thick) for different parameters of the winds produced by the massive stars: classical WR type star (dot-dashed), $\eta$ Carinae type star (solid) and WR star in WR 20a binary system (dashed). The dotted line shows the adiabatic energy losses of hadrons without their collisional energy losses.}
\label{fig3}
\end{figure}
\section{Gamma-rays from protons in the wind cavity}

In this section we calculate the spectra of $\gamma$-rays and neutrinos produced within the wind cavity.
As shown above, primary nuclei are completely fragmented on separate nucleons (protons and neutrons) in collisions with the matter and radiation already within the binary systems. These nucleons are accelerated with the power law spectrum. Protons from disintegration of nuclei are captured by the magnetic field of the stellar winds. They are advected from the binary system with the velocity of the wind. They produce
high energy radiation in collisions with the matter of the stellar wind during their propagation within the wind cavity.
These $\gamma$-rays are produced isotropically since the production mechanism (hadronic collisions) is independent on the radiation field of the massive star.
On the other hand, neutrons, extracted from nuclei, move rectilinearly through the wind cavity. They decay at distances from the binary system  which depend on their Lorentz factors. Secondary protons, from decaying neutrons, are also captured in the stellar wind. All these secondary protons suffer collisional and adiabatic energy loses due to the expansion of the wind.
In order to obtain the $\gamma$-ray spectra, escaping from the wind cavity region to the observer, we have to consider possible absorption of $\gamma$-rays produced in hadronic collisions in the radiation field of the massive stars. 
Therefore, we calculate the optical depths for $\gamma$-rays. These absorption effects has to be taken into account when calculating the $\gamma$-ray spectra produced by hadrons not far from the binary system.

\subsection{Absorption of $\gamma$-rays close to the massive star}

\begin{figure*}
\vskip 6.5truecm
\includegraphics{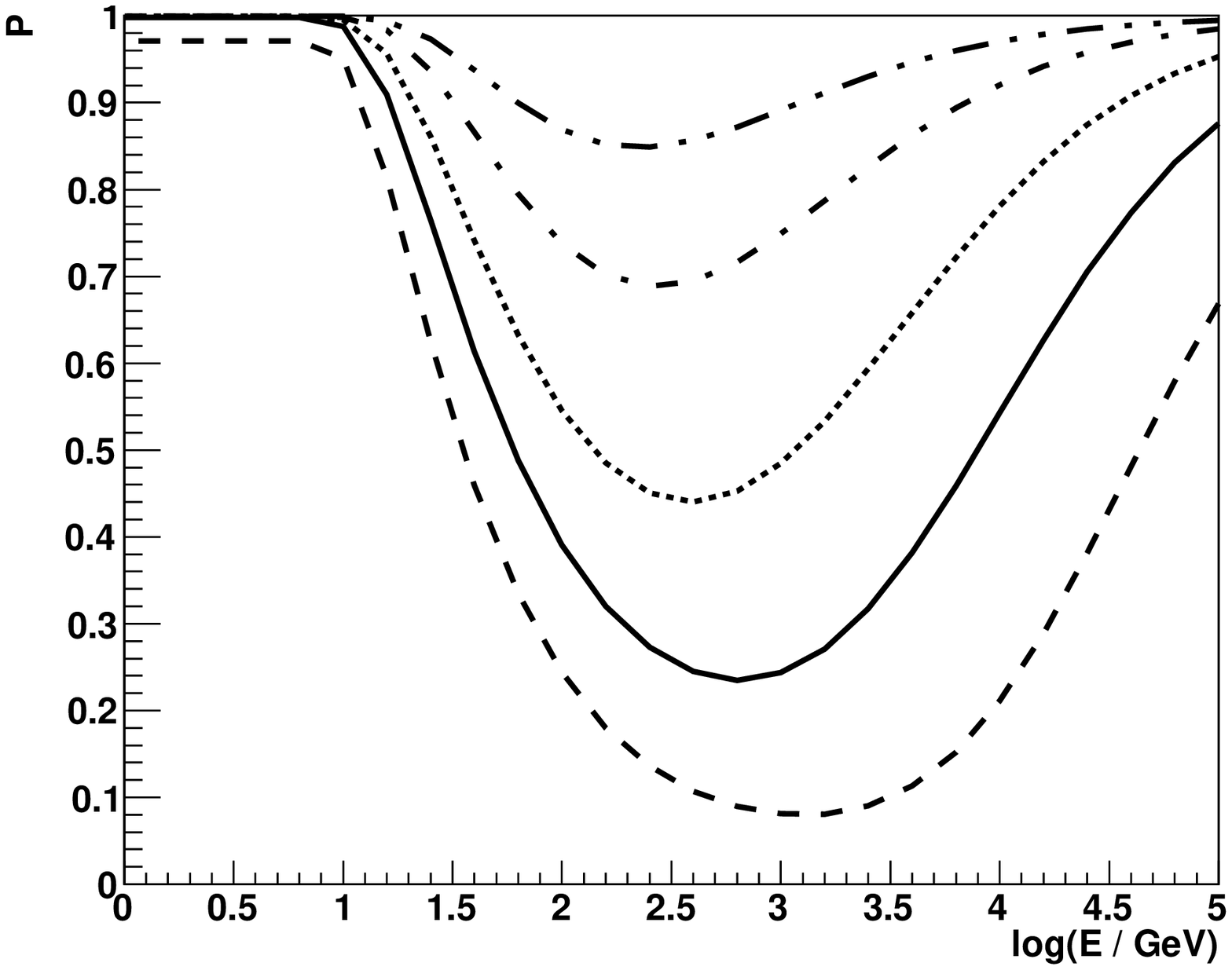}
\includegraphics{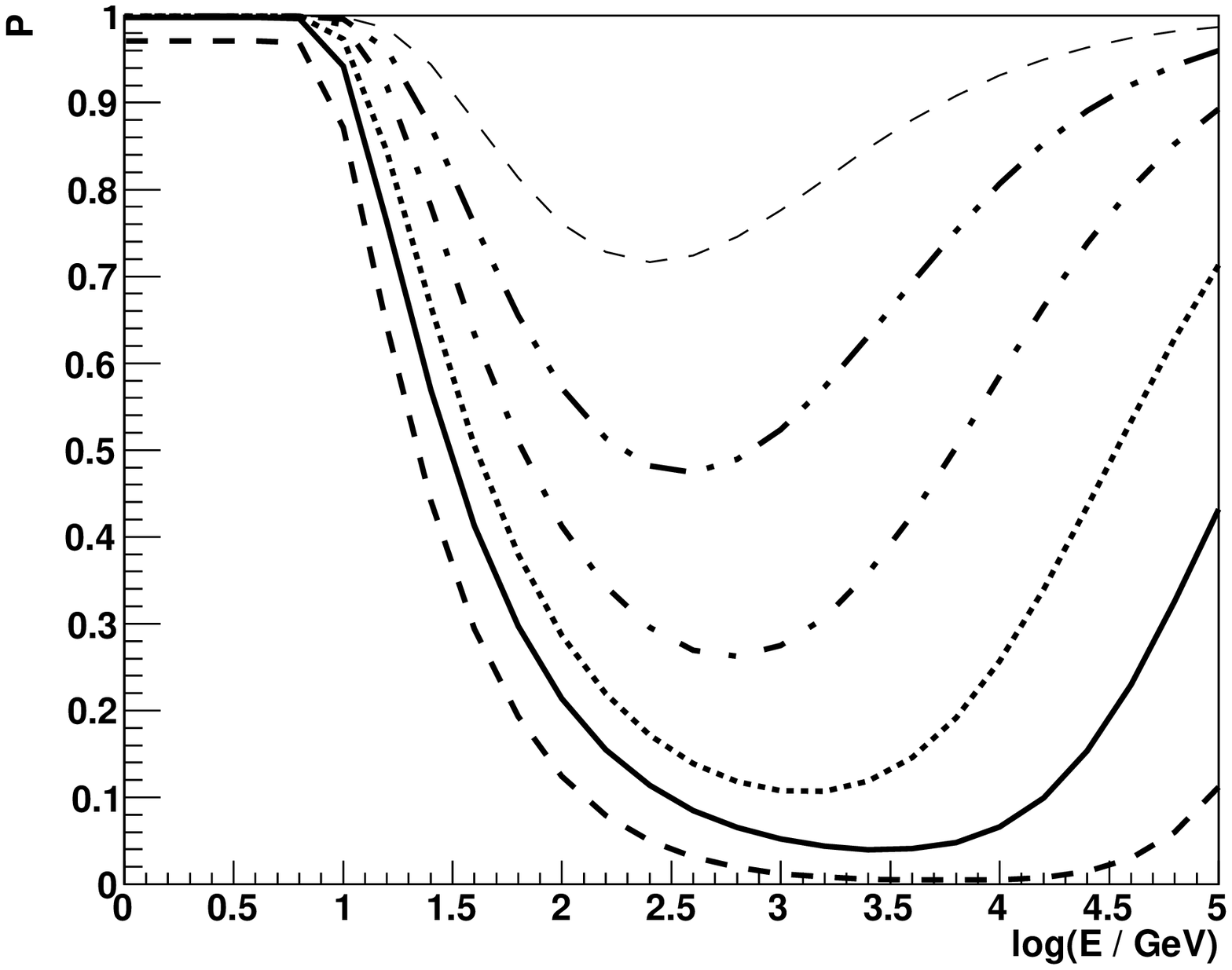}
\caption{The average $\gamma$-ray reduction factors, $P$ (see Eq.~10), due to absorption of $\gamma$-rays in the stellar radiation ($\gamma + \gamma\rightarrow e^\pm$), as a function of the energy of $\gamma$-ray photon. The results are shown  for the massive star in WR 20a (on the left) and Eta Carinae (on the right) binary systems. The reduction factors are calculated assuming isotropic injection of $\gamma$-rays at a specific distance from the massive star in units of the stellar radius equal to r = 3 (dashed), 10 (solid), 30 (dotted), 100 (dot-dashed), 300 (dot-dot-dashed), and $10^3$  (thin dashed).}
\label{fig4}
\end{figure*}

Significant amount of $\gamma$-rays produced in hadronic collisions (as described above) originate close enough to the massive star that their absorption in the stellar radiation can play important role. The details of the absorption process are
determined by the geometry of the binary system (production place of $\gamma$-rays in respect to the star) and the parameters of the massive star. Its importance in the case of binary systems without precise orbital parameters can be approximately evaluated by calculating the average reduction factors for $\gamma$-ray photons, i.e. the probability of escape averaged over the whole range of possible propagation angles in respect to the direction towards the star. We calculate the reduction factors for $\gamma$-rays injected isotropically at the distance R from the massive star from,
\begin{eqnarray}
P = 0.5\int_{-1}^{1}e^{-\tau_{\gamma\gamma}(R,\mu)}d\mu,
\label{eq10}
\end{eqnarray} 
\noindent
where $\tau_{\gamma\gamma}(R,\mu)$ is the angle dependent optical depth for the $\gamma$-ray photon injected at the distance R, $\mu = \cos\theta$, and at the angle $\theta$ in respect to the direction defined by the injection place and the center of the star (see e.g. Bednarek~1997, B\"ottcher \& Dermer~2005, Dubus~2006, Zdziarski et al.~2014).
These optical depths can be easily re-scaled for stars with different parameters following simple prescription given by Eq.~2 in Bednarek \& Pabich~(2010).
The absorption process concerns mainly $\gamma$-rays with energies above $E_\gamma^{\rm min} = m_{\rm e}/3k_{\rm B}T\sim 20/T_5$ GeV. We calculate the optical depths for $\gamma$-rays as a function of their energies and the distance from the massive star.
As we noted above, the exact values of the optical depths depend on the propagation angle of $\gamma$-rays in respect to the direction towards the star. Therefore, we show the reduction factor values averaged over the injection angles assuming isotropic injection of $\gamma$-rays from a point source at a specific distance from the star. The results of these example calculations are shown in Fig.~4 for the WR 20a and Eta Carinae massive stars.
It is clear that $\gamma$-rays with energies above several GeV are efficiently absorbed even if they are produced at relatively large distances from the binary system.

\subsection{Protons from disintegrated nuclei}

We consider the production of radiation by protons from nuclei accelerated within the binary system. These protons are advected outside the binary system with the velocity of the stellar wind. The number of hadrons which interact during the advection process with the stellar wind, on the range of distances from the binary system between $R_{\rm BS}$ (considered as the injection place) and R, is determined by the factor $[1 - exp({-\tau_{pp}})]$, where $\tau_{\rm pp} = A(1/R_{\rm BS} - 1/R)$ (see Eq.~8), and  $A = {\dot M}c\sigma_{\rm pp}/(4\pi v_{\rm w}^2)\approx 2.9\times 10^{12}{\dot M}_{-5}/v_3^2$ cm.
The spectrum of hadrons, which interact at the distance $R$ from the binary system, is
\begin{eqnarray}
N_{\rm h} = {{dN_{\rm h}(\gamma_{\rm h},R)}\over{d\gamma_{\rm h}dRdt}} = 
{\dot N}_{\rm h} J {{A}\over{R^2}}
e^{A(1/R - 1/R_{\rm BS})},
\label{eq11}
\end{eqnarray}
\noindent
where ${\dot N}_{\rm h} = dN(\gamma_{\rm h})/d\gamma_{\rm h}dt$ is the injection rate of hadrons with the Lorentz factors $\gamma_{\rm h}$ from the binary system, $J = d\gamma_{\rm h}(R_{\rm BS})/d\gamma_{\rm h}(R) = R/(R_{\rm BS}K^{\tau_{\rm hp}})$ is the Jacobian of transformation of hadron energy (see Eq.~\ref{eq9}).  

In order to calculate the $\gamma$-ray and neutrino spectra produced by protons, we have to integrate the above injection rate of protons (Eq.~\ref{eq11}) over the whole dimension of the wind cavity and over the spectrum protons (given by Eq.~11),
\begin{eqnarray}
{{dN_{\gamma,\nu}}\over{dE_{\gamma,\nu}dt}} =
\int_{\rm R_{\rm BS}}^{\rm R_{\rm c}}
\int_{\gamma_{\rm min}}^{\gamma_{\rm max}}
N_{\rm h}
{{dN_{\gamma,\nu}(\gamma_{\rm h})}\over{dE_{\gamma,\nu}}}e^{-\tau_{\gamma\gamma}}
d\gamma_{\rm h}dR 
\label{eq12}
\end{eqnarray}
\noindent
The spectra of $\gamma$-rays and neutrinos are calculated by applying the scaling break model for hadronic collisions developed by Wdowczyk \& Wolfendale (1987), which is suitable in the considered energy range of relativistic hadrons.
The distribution of protons within the wind cavity, $N_{\rm h}$, is given by Eq.~\ref{eq11}.
The spectra are calculated for the range of energies of hadrons $\gamma_{\rm min} = 10$ and $\gamma_{\rm max}$ as reported in Table~1. $E_\gamma$ and $E_{\nu}$ are the energies of $\gamma$-ray photons and neutrinos, respectively. The example spectra of $\gamma$-rays, calculated for two binary systems (WR 20a and Eta Carinae), are
shown in Fig.~5. We show the $\gamma$-ray spectra with (thick curves) and without (thin curves) absorption effects in the radiation of the massive stars. As expected, the absorption of $\gamma$-rays produced by protons, extracted from nuclei within the inner part of the wind cavity, is very strong.

\begin{figure*}
\vskip 6.5truecm
\includegraphics{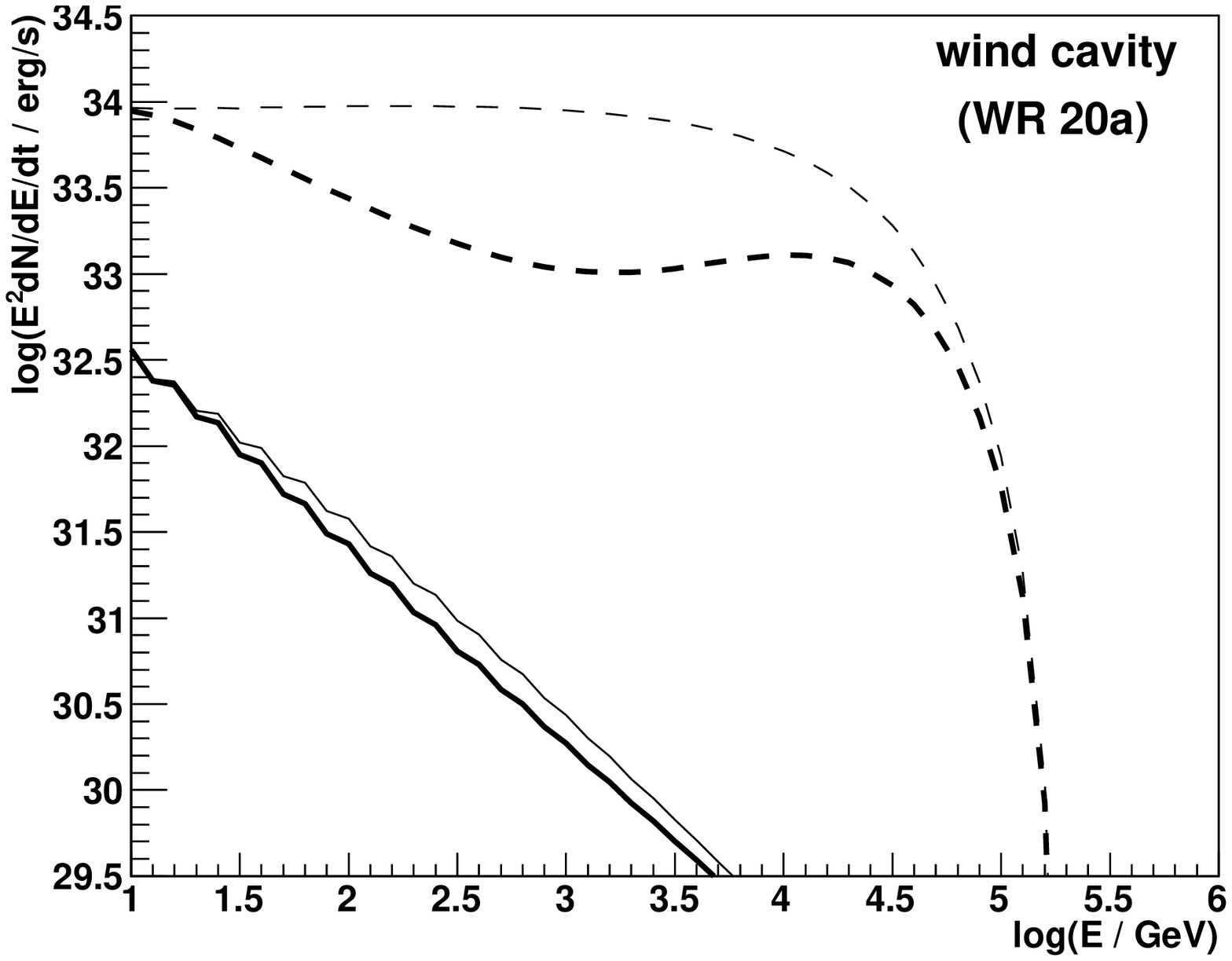}
\includegraphics{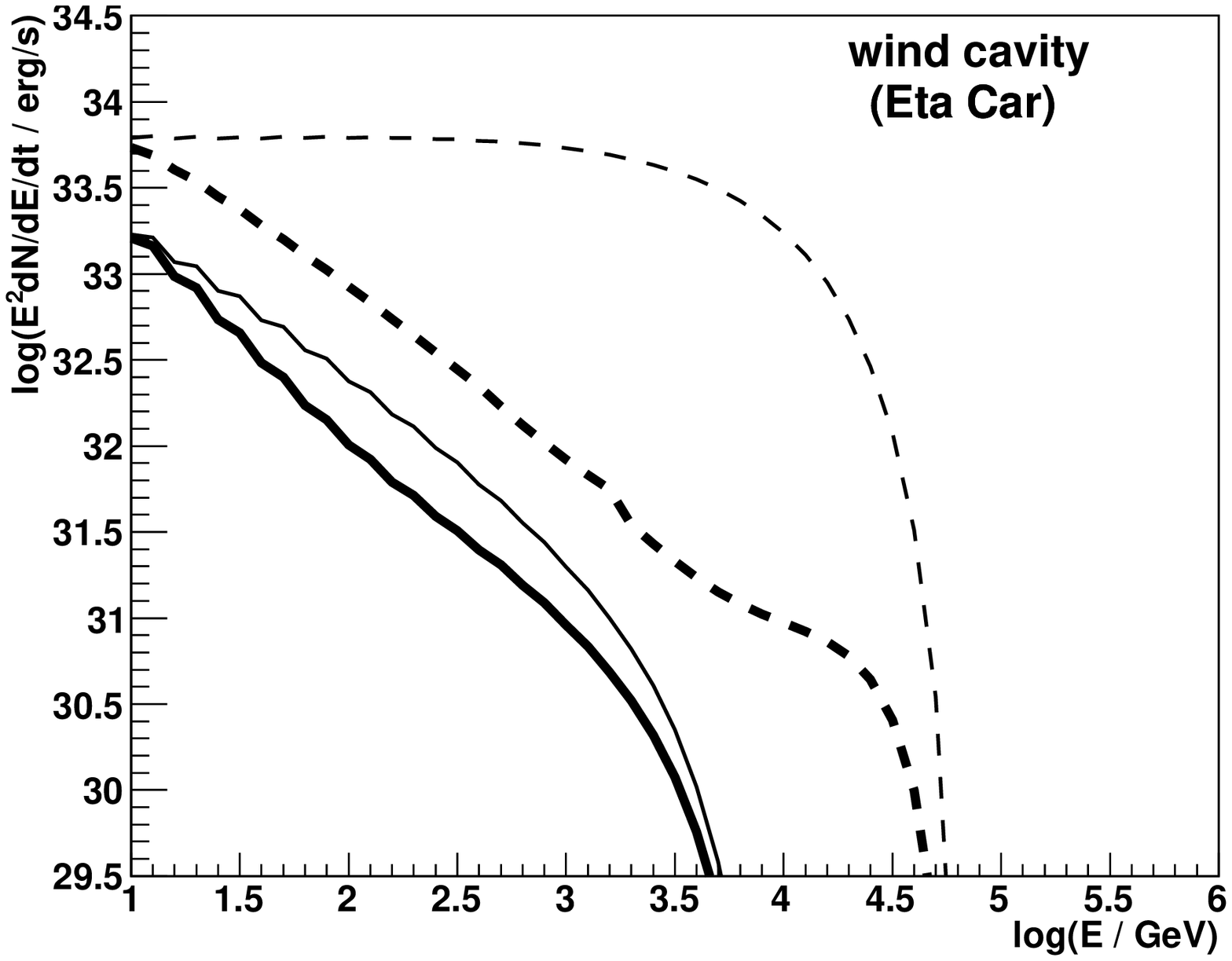}
\caption{Gamma-ray spectra produced by protons, from disintegrated nuclei, which interact within the matter of the stellar winds within the wind cavity (dashed curves) and from protons, appearing in the wind cavity as a decay products of neutrons (solid curves). Two example binary systems are considered WR 20a (on the left) and Eta Carinae (on the right). The thick curves show the gamma-ray spectra with included absorption effects in the stellar radiation and the thin curves show the spectra without any $\gamma$-ray absorption. 
The $\gamma$-ray spectra have been calculated for the spectra of primary nuclei which have been normalized to the stellar wind powers (see Table~1), with the normalization coefficients equal to $\eta = 10^{-2}$.}
\label{fig5}
\end{figure*}

\subsection{Protons from decaying neutrons}

\begin{figure*}
\vskip 6.5truecm
\includegraphics{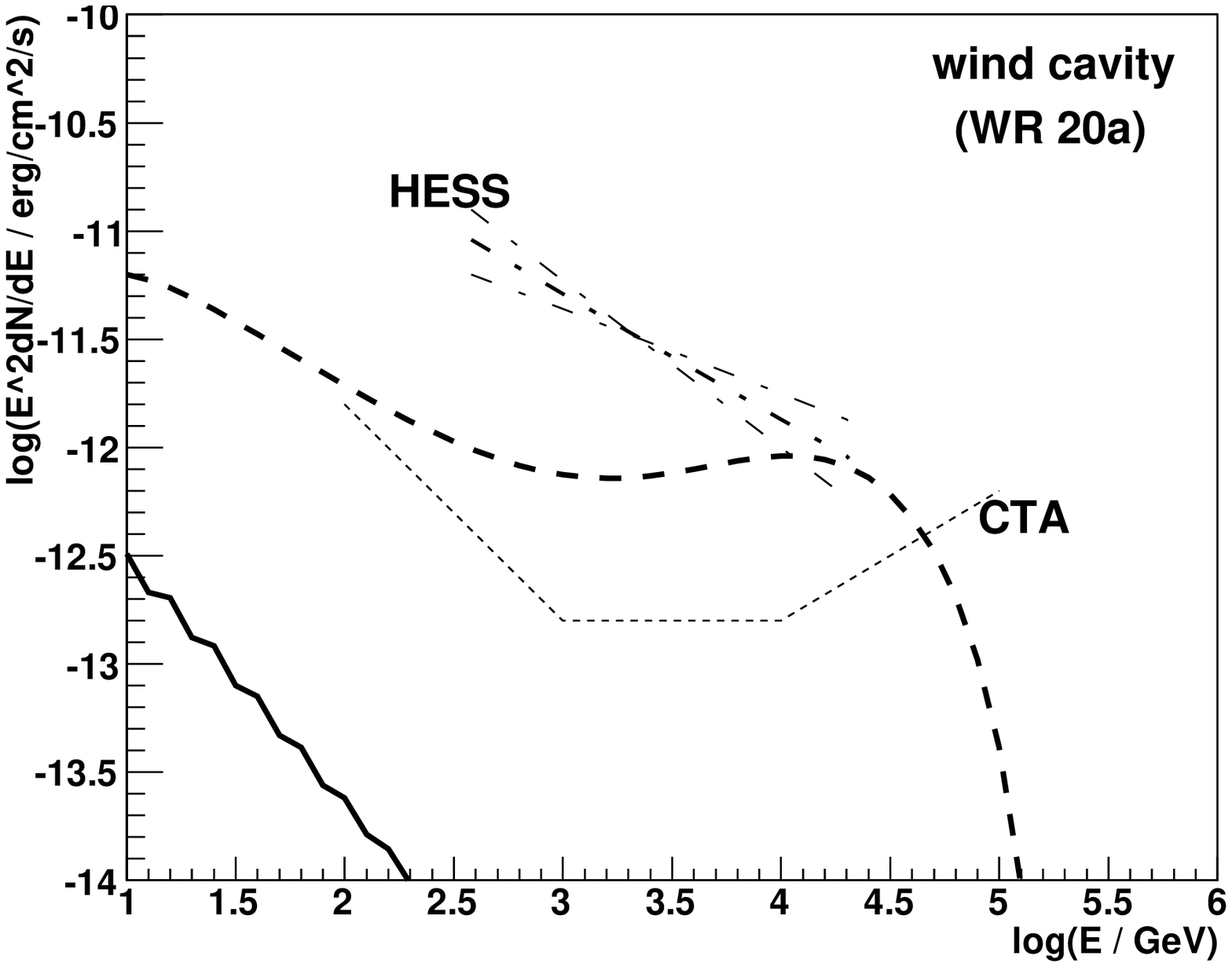}
\includegraphics{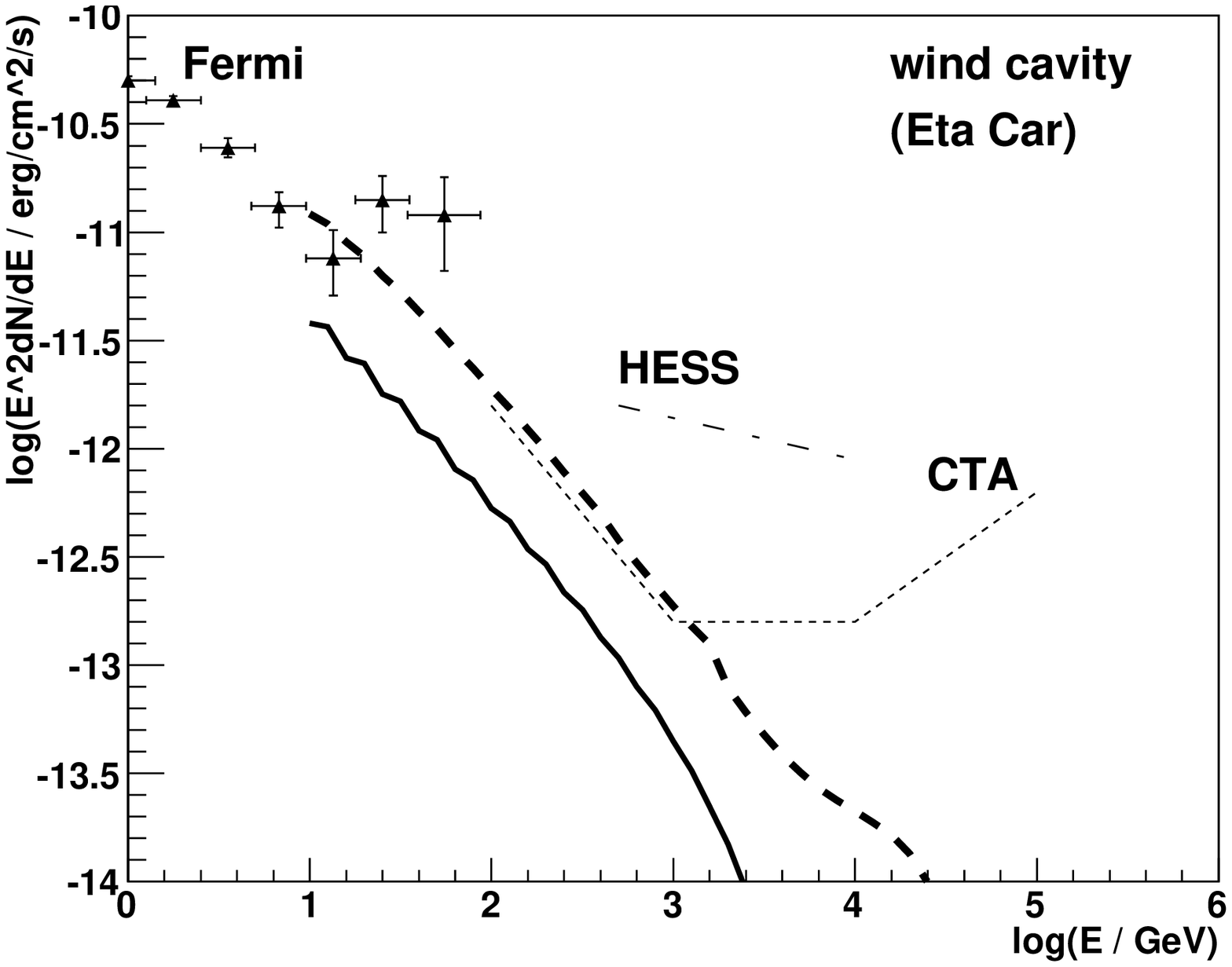}
\caption{Gamma-ray spectra produced by protons from disintegrated nuclei (dashed curves), and by protons (solid curves) decaying neutrons (solid curves). These protons interact with the matter within the wind cavity in the case of two example binary systems the WR 20a (on the left) and Eta Carinae (on the right). The absorption effects of $\gamma$-rays in the stellar radiation are included. 
The $\gamma$-ray spectrum observed by HESS from the direction of WR 20a binary system is shown by the dot-dashed line (Abramowski et al.~2011). In the case of Eta Carinae binary system, the Fermi-LAT spectrum  is shown by triangles (Abdo et al.~2009) and the HESS upper limits (Abramowski et al.~2012b) are shown by the thin dot-dashed line. The $\gamma$-ray spectra have been calculated for the spectra of primary nuclei which have been normalized to the stellar wind powers (see Table~1). The normalization coefficients are equal to $\eta = 3\times 10^{-3}$, in order to be consistent with the $\gamma$-ray flux reported by HESS from the open cluster Westerlund 2 (see thin dot-dashed line), and $\eta = 10^{-2}$, in order to be consistent with the Fermi observations of Eta Carinae below 10 GeV. The broken thin dotted line show the level of sensitivity of the CTA (Bernl\"ohr et al.~2012).}
\label{fig6}
\end{figure*}

Neutrons, extracted from nuclei within the binary system, propagate along the straight lines and gradually decay into protons in the dense medium surrounding the binary system. Depending on the Lorentz factors of neutrons, these protons appear within the wind cavity or outside the wind cavity (i.e. within the open cluster). In this subsection, we calculate the spectra of protons from neutrons decaying within the wind cavity. These spectra are used for the calculation of $\gamma$-ray and neutrino spectra produced within the wind cavity.

The optical depth for neutrons, which move along straight lines through the wind cavity, on the interaction with the matter of the wind is equal to $\tau_{\rm np} = \int_{R_{\rm BS}}^{R_{\rm c}}\sigma_{\rm pp}n_{\rm w}dR$, where $n_{\rm w}$ is defined by Eq.~1. It is a factor of $c/v_{\rm w}$ smaller than the optical depth for protons given in Table~1.
Therefore, the interaction rate of neutrons moving along the straight lines through the wind cavity can be neglected for the considered range of the parameters characterizing the binary systems. 
The neutron decay rate at the distance, $R'$, from the binary system (equal to the rate of creation of protons from their decay) is given by, 
\begin{eqnarray}
{{dN_{\rm p}(\gamma_{\rm p},R')}\over{d\gamma_{\rm p}dR'dt}} = 
{{{\dot N}_{\rm n}}\over{c\gamma_{\rm n}\tau_{\rm n}}}e^{-R'/(c\gamma_{\rm n}\tau_{\rm n})},
\label{eq13}
\end{eqnarray}
\noindent
where ${\dot N}_{\rm n} = dN/d\gamma_{\rm n}dt$ is the neutron injection rate from the binary system (equal to the acceleration rate of nuclei), $\tau_{\rm n} = 900$ s is the neutron decay time,
and the Lorentz factor of secondary protons ($\gamma_{\rm p}$) is assumed to be equal to the Lorentz factor of neutrons ($\gamma_{\rm n}$). Protons, from decaying neutrons, are confined by the magnetic field of the stellar wind. Therefore, they suffer adiabatic and collisional energy losses as considered for protons from fragmentation of nuclei in the previous section. 
The collision rate of protons at the distance, $R > R'$, from the binary system, is given by the formula similar to that given by Eq.~\ref{eq11},
\begin{eqnarray}
{{dN_{\rm p}(\gamma_{\rm p},R',R)}\over{d\gamma_{\rm p}dR'dRdt}} = 
{{dN_{\rm p}}\over{d\gamma_{\rm p}dR'dt}}
{{A}\over{R^2}} J e^{A(1/R' - 1/R)} ,
\label{eq14}
\end{eqnarray}
\noindent
where $J = d\gamma_{\rm p}(R')/d\gamma_{\rm p}(R) = R/(R'k^{\tau_{\rm pp}})$, and $\tau_{\rm pp} \approx 
2.9\times 10^{12}({\dot M}_{-5}/v_3^2)(1/R'- 1/R) = A (1/R'- 1/R)$.

The $\gamma$-ray and neutrino spectra, produced by these secondary protons in the wind cavity, can be calculated by integrating the above formula over the spectrum of protons, their creation distance in the cavity measured from binary system, $R'$, and their interaction distance $R$,
\begin{eqnarray}
{{dN_{\gamma,\nu}}\over{dE_{\gamma,\nu}dt}}  & = &
\int_{\rm R_{\rm BS}}^{\rm R_{\rm c}}dR\int_{\rm R}^{R_{\rm c}}dR'
\int_{\gamma_{\rm min}}^{\gamma_{\rm max}}d\gamma_{\rm p} \cr 
&  & {{dN_{\rm p}}\over{d\gamma_{\rm p}dRdR'dt}}
{{dN_{\gamma,\nu}(\gamma_{\rm p}(R))}\over{dE_{\gamma,\nu}}}, 
\label{eq15}
\end{eqnarray}
\noindent
where $dN_{\rm p}/dRdR'd\gamma_{\rm p}dt$ is given by Eq.~\ref{eq14}, and the spectra of $\gamma$-rays and neutrinos are calculated as described below Eq.~12.

We have performed calculations of the $\gamma$-ray spectra produced by protons from neutrons decaying within the wind cavity for the parameters of two considered binary systems (see thick solid curves in Fig.~5). These spectra are clearly below the $\gamma$-ray spectra produced by protons directly extracted from nuclei. This effect is due to
the appearance of protons (from decaying neutrons) at a relatively large distances from the binary system where the density of the stellar wind is already low. On the other hand, these $\gamma$-ray spectra show much smaller absorption effects in the stellar radiation due to their production at larger distances from the massive star.

\subsection{Confrontation with observations of specific binary systems}

The $\gamma$-ray spectra produced in the wind cavity by two considered above populations of protons are compared with the available observations of the two considered binary systems in the GeV-TeV energy range and with the sensitivity of the planned CTA. As mentioned in the Introduction, the TeV $\gamma$-ray source has been reported in the direction of the open cluster Westerlund 2
which contains binary system WR 20a (Aharonian et al.~2007, Abramowski et al.~2011). The $\gamma$-ray spectrum of the source in the direction of this open cluster has the spectral index 2.58 in the energy range $\sim$1-10 TeV.
The nature of this source is at present unknown. It is supposed that this emission can be related to the binary system WR 20a, the pulsar wind nebula (PWNe) around PSR J1022-5746 or maybe also to the dense molecular clouds present in this open cluster. We compare the $\gamma$-ray spectrum expected from the wind cavity around the binary system WR 20a with the above mentioned observations. The results are shown in Fig.~6. In order to be consistent with the TeV observations, reasonable value for the energy conversion efficiency from the stellar wind to relativistic nuclei is required ($\sim 5\times 10^{-3}$). According to our calculations, $\gamma$-rays produced in the wind cavity should be mainly responsible for a part of this emission at the highest observed energies, i.e. $\sim$ 10 TeV.
As we show below, the lower energy part of the $\gamma$-ray emission from Westerlund 2 could be either produced by protons which escape from the wind cavity into dense regions of the open cluster, as considered in Sect.~7, or it comes from the other sources present within the Westerlund 2 (e.g. other massive binary systems or PWNe). 

We also compare our calculations with the observations of the binary system Eta Carinae and the surrounding dense complex Carina Nebula. The source in this direction has been detected in the GeV energies by the Agile (Tavani et al.~2009) and Fermi telescopes (Abdo et al.~2009). $\gamma$-ray emission from this source shows two components spectrum, first one extending to a few GeV and the second one extending up to $\sim$ 100 GeV (Farnier et al.~2011). The highest energy component show evidences of variability with the orbital period of the binary system (Walter \& Farnier~2011, Reitberger et al.~2012). This component is expected to be produced within the binary system. 
On the other hand, the lower component seems to be steady. We compare the $\gamma$-ray spectrum expected in terms of our model from the wind cavity around the binary system in Fig.~6. The spectrum is normalized to the observed lower energy component extending to a few GeV. Since the $\gamma$-ray emission produced within the wind cavity is strongly absorbed in the stellar radiation, the $\gamma$-ray spectrum expected in our model in the TeV energy range is steep (spectral index close to -4, see Fig.~6). Therefore, this emission is clearly below the present upper limit reported by the HESS Collaboration in the TeV energy range (Abramowski et al.~2012b). Moreover, due to the steepness of this spectrum, future detection of this emission component by the CTA is rather problematic (Fig.~6).

\begin{figure*}
\vskip 6.5truecm
\includegraphics{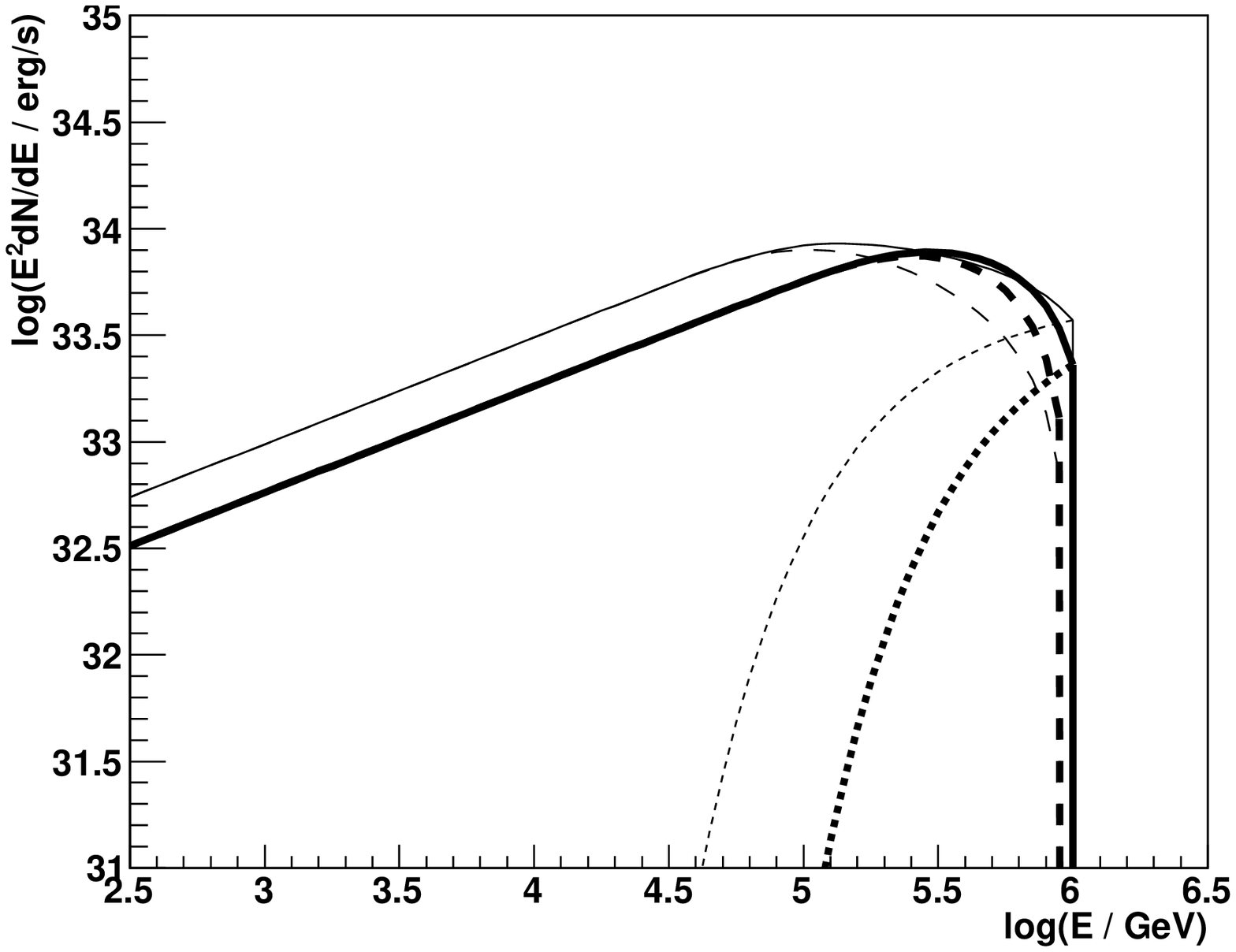}
\includegraphics{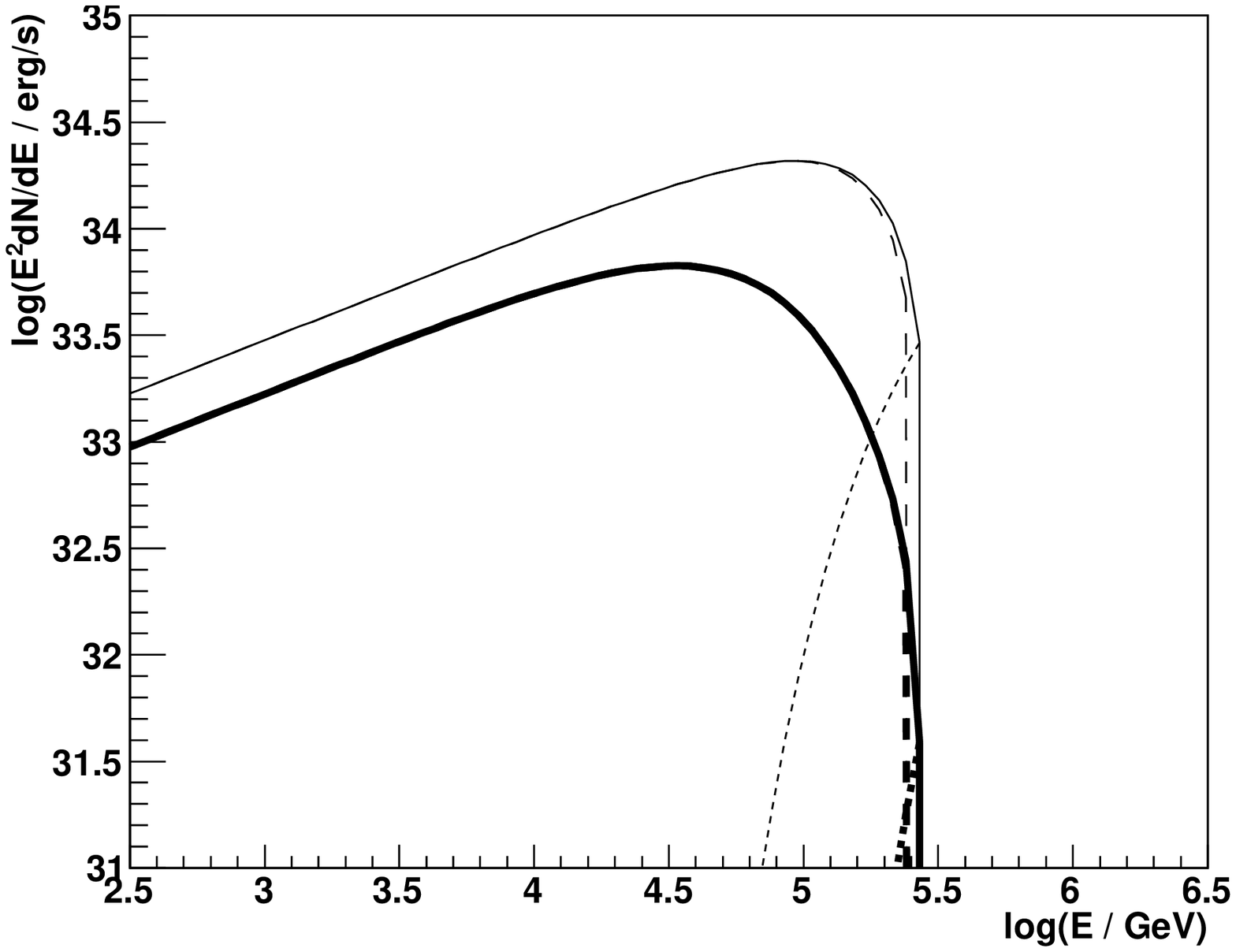}
\caption{Differential spectra of protons (Spectral Energy Distribution - SED) from neutrons decaying within the wind cavity but advected from it to the open cluster (dashed curves) and from neutrons decaying outside the wind cavity but inside the open cluster (dotted curves). Total spectra of protons within the open cluster are shown by the solid curves. 
Neutrons are extracted from nuclei in collisions with the matter of the stellar wind 
in the case of WR 20a binary system (on the left) and Eta Carinae binary system (on the right).
Density of matter in the open cluster determines the radius of the wind cavity (see Eq.~2), $n_{\rm oc} = 10$ cm$^{-3}$ (thick curves) and $n_{\rm oc} = 100$ cm$^{-3}$ (thin). The power in relativistic nuclei is normalized to $1\%$ of the stellar wind power (given in Table~1).}
\label{fig7}
\end{figure*}

\section{Gamma-rays from protons in the open cluster}

In this section we calculate the $\gamma$-ray (and neutrino) spectra produced by protons which originate as a decay products of neutrons within the wind cavity, but they are advected with the stellar wind into the surrounding open cluster. We also calculate the $\gamma$-ray emission from relativistic protons which originate from the most energetic neutrons decaying directly outside the wind cavity, i.e. within the open cluster.
It is assumed that the open cluster has the basic parameters of the order of, the density of matter 10 cm$^{-3}$, the magnetic field strength $10^{-4}$ G, and the radius 20 pc. As we show below, the combination of these parameters determines the escape of protons from the open cluster and, as a consequence the $\gamma$-ray production rate and their spectra in collisions with the matter of the open cluster.

\subsection{Protons advected outside the wind cavity}

A part of protons, which is advected with the stellar wind, can arrive up to the wind cavity border. They can be injected into 
the dense open cluster surrounding the binary system. However, only protons from decaying neutrons can still have large energies allowing them to produce the TeV $\gamma$-rays and neutrinos. Protons, from direct disintegration of nuclei, appear close to the binary system and therefore suffer huge adiabatic energy losses. The spectrum of protons injected from the wind cavity (with the radius $R_{\rm c}$) into the open cluster can be 
calculated from,
\begin{eqnarray}
{{dN_{\rm p}(\gamma_{\rm p})}\over{d\gamma_{\rm p}dt}} =
\int_{\rm R_{\rm BS}}^{\rm R_{\rm c}} 
{{dN_{\rm p}}\over{d\gamma_{\rm p}dR'dt}}JdR', 
\label{eq16}
\end{eqnarray}
\noindent
where $dN_{\rm p}/d\gamma_{\rm p}dR'dt$ is given by Eq.~13
and the Jacobian is $J = d\gamma_{\rm p}(R')/d\gamma_{\rm p}(R_{\rm c})$.

We calculate the spectra of protons advected into the open cluster for two considered binary systems (WR 20a and Eta Carinae).
Two different values for the density of matter within the open cluster, which determine the outer radius of the wind cavity, are assumed.
The results are shown in Fig.~7 (see dashed curves). As expected, for denser matter within the open cluster (corresponding to smaller radius of the wind cavity) the spectra have larger intensities and the maximum in the spectra are shifted to lower energies. These effects are due to the lower adiabatic energy losses in the case of wind cavities with smaller radii.

In order to calculate the $\gamma$-ray and neutrino spectra produced by these protons in the open cluster, we have to estimate their residence time within the open cluster, due to the diffusion process, and compare it with the collisional time scale.
We assume that protons diffuse in the turbulent medium of the open cluster with the rate well determined by the Bohm diffusion coefficient,
$D_{\rm B} = R_{\rm L}c/3\approx 3\times 10^{26}\gamma_6/B_{-4}$ cm$^2$s$^{-1}$, where the magnetic field strength within the open cluster is $B_{\rm oc} = 10^{-4}B_{-4}$ G. Then, the average diffusion time scale of protons within the open cluster, with the characteristic dimension $R_{\rm oc} = 20R_{20}$ pc, is estimated on,
$\tau_{\rm dif} = R_{\rm oc}^2/2D_{\rm B}\approx 7\times 10^{12}R_{20}^2B_{-4}/\gamma_6$ s. 

The collision time of these protons can be estimated from
$t_{\rm p}^{\rm out} = (cn_{\rm oc}\sigma_{\rm pp})^{-1}\approx
10^6/n_{30}$ yrs, where the density of surrounding matter is $n_{\rm oc} = 30n_{30}$ cm$^{-3}$. The interaction process of these protons within the open cluster depends on the lifetime of the massive binary system and the diffusion time scale of protons from the open cluster. The time scale for  collisions of protons with the matter is, 
$\tau_{pp} = (c\sigma_{\rm pp}n_{\rm cl})^{-1}\approx 10^{14}/n_{10}$ s.

By comparing the diffusion time scale with collision time scale, we estimate critical energy of relativistic protons, 
\begin{eqnarray}
\gamma_{\rm p} < \gamma_{\rm p}^{\rm int} = 7\times 10^4R^2_{20}B_{-4}n_{10} 
\label{eq17}
\end{eqnarray}
\noindent
below which they can interact efficiently within the open cluster. We assume that protons with such energies reaches steady state equilibrium
in which the rate of proton injection equals the rate of proton interaction. We calculate the $\gamma$-ray and neutrino spectra produced by protons with 
the spectra shown in Fig.~7 in the range limited by Eq.~17.

Note that the winds from the massive stars in the binary system do not change significantly the distribution of the matter
within the open cluster since the mass swept-up by the stellar wind is 
$M_{\rm sur} = 4R_{\rm c}^3n_{\rm oc}/3 \approx 90({\dot M}_{-5}v_3n_{10}/T_4)^{1/2}$ M$_\odot$.
This is only a small amount of the total mass of the open cluster which is typically expected in the range $10^3$ M$_\odot$ to 
a few $10^4$ M$_{\odot}$.

The $\gamma$-ray and neutrino spectra produced by protons, advected from the wind cavity, are calculated by integration of the above derived proton spectrum, 
\begin{eqnarray}
{{dN_{\gamma,\nu}}\over{dE_{\gamma,\nu}dt}} =
\int_{\gamma_{\rm min}}^{\gamma_{\rm p}^{\rm int}}
{{dN_{\rm p}(\gamma_{\rm p})}\over{d\gamma_{\rm p}dt}}
{{dN_{\gamma,\nu}(\gamma_{\rm h})}\over{dE_{\gamma,\nu}}}
d\gamma_{\rm p}
\label{eq18}
\end{eqnarray}
\noindent
where $dN_{\rm p}(\gamma_{\rm p})/d\gamma_{\rm p}dt$ is given by Eq.~(16).

\subsection{Protons from neutrons decaying outside the wind cavity}

The most energetic neutrons, extracted from nuclei within the binary system, can decay directly outside the wind cavity,
since their propagation distance $L_{\rm n} = \gamma_{\rm n}\tau_{\rm n} c$ becomes comparable to the radius of the wind cavity. 
The spectrum of protons from decay of these neutrons is given by,
\begin{eqnarray}
{{dN_{\rm p,cl}(\gamma_{\rm p})}\over{d\gamma_{\rm p}dt}} = 
{{dN_{\rm n,cl}(\gamma_{\rm n})}\over{d\gamma_{\rm n}dt}} = {\dot N}_{\rm n}e^{{-R_{\rm c}}\over{(\gamma_{\rm n}\tau_{\rm n}c)}},
\label{eq19}
\end{eqnarray}
\noindent
where ${\dot N}_{\rm n} = dN(\gamma_{\rm n})/d\gamma_{\rm n}dt$ is the spectrum of neutrons extracted from nuclei, 
$R_{\rm cav}$ is the radius of the wind cavity (see Eq.~2). This neutron spectrum has exactly the same shape as the spectrum of accelerated nuclei, i.e. the power law type with the spectral index equal to 2 (see Section.~3). It is assumed that Lorentz factors of neutrons are equal to Lorentz factors of parent nuclei.
Proton spectra, from neutrons decaying inside the open cluster (i.e. outside the wind cavity), are shown for both considered binary systems in Fig.~7 (dotted curves). Relative contribution of protons from these neutrons, in respect to protons advected from the wind cavity, depends on the parameters defining the acceleration scenario, the wind of the companion stars and the parameters of the open cluster.
The largest contribution is expected in the case of binary systems containing massive stars with strong surface magnetic field (large Lorentz factors of nuclei) and relatively weaker stellar winds (smaller radius of the wind cavity). Therefore, more neutrons should decay directly within the open cluster (outside the wind cavity) in the case of the WR type stars than in the case of the most massive stars of the Eta Carinae type which have very large mass loss rates but weaker surface magnetic fields.

The $\gamma$-ray and neutrino spectra, produced by these protons, are calculated by integrating formula given by Eq.~18.

\begin{figure*}
\vskip 6.5truecm
\includegraphics{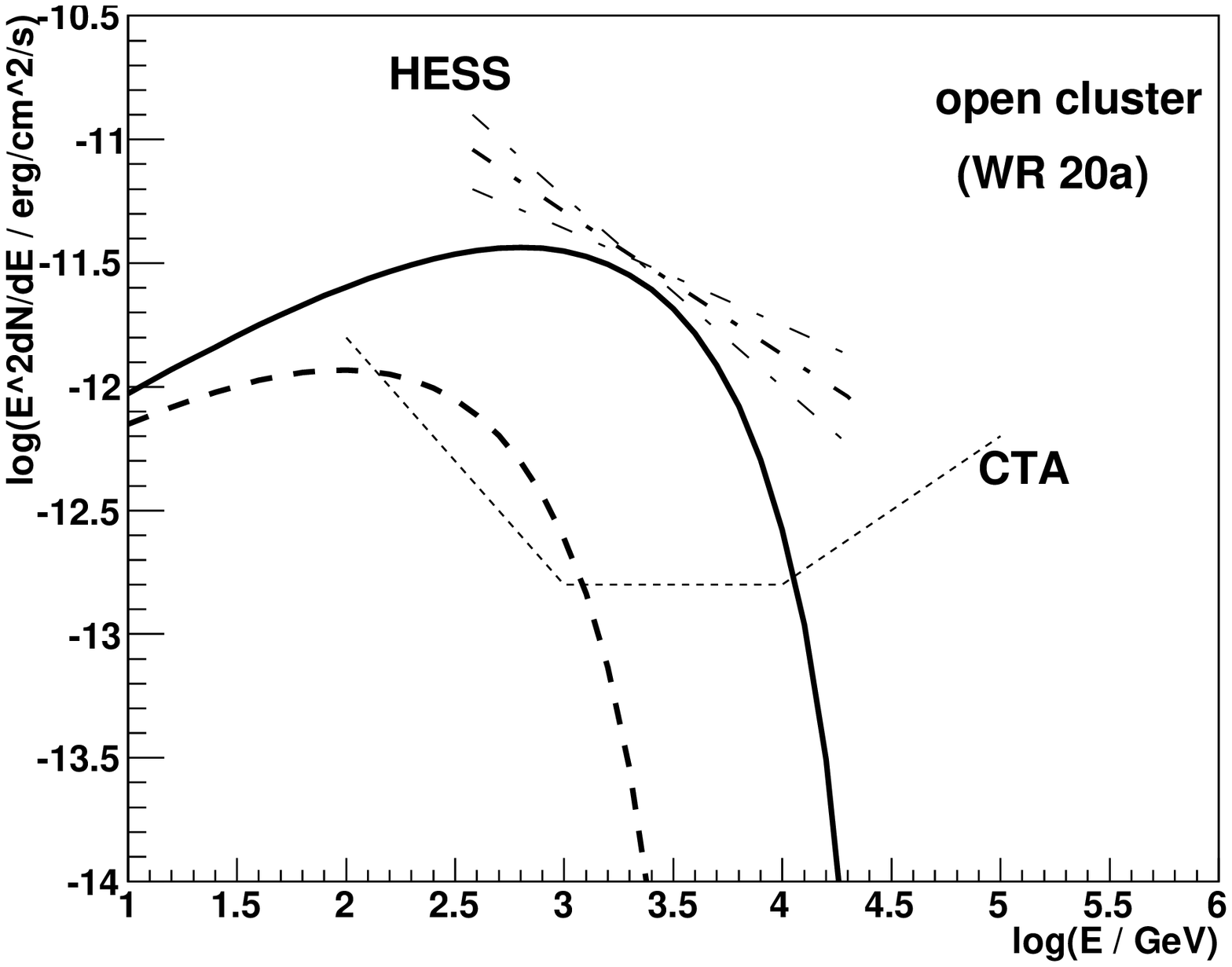}
\includegraphics{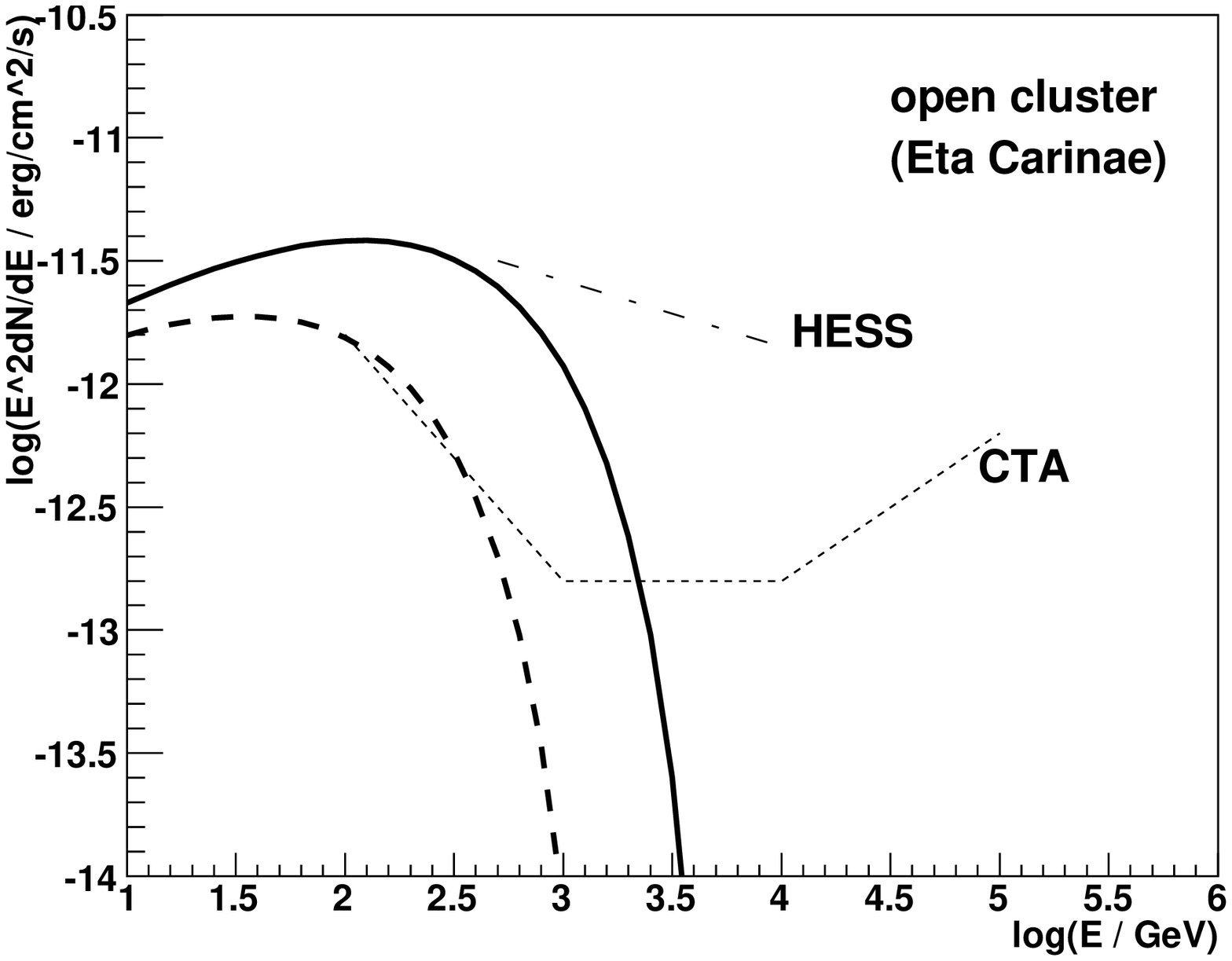}
\caption{Gamma-ray spectra produced by protons (from decaying neutrons) in the dense cluster around WR 20a and Eta Carinae binary systems.
The escape of protons from the open cluster is described by assuming the Bohm diffusion prescription and the parameter $R_{20}^2B_{-4}n_{10} = 1.5$ (solid curve) and 0.15 (dashed) for WR 20a binary, and 0.2 (solid) and 0.04 (dashed) for Eta Carinae binary. The HESS spectrum observed from the direction of the open cluster around WR 20a is marked by the thin dot-dashed line and the CTA sensitivity is marked by the thin dotted curve. The efficiency of acceleration of nuclei is equal to $\eta = 5\times 10^{-3}$ for WR 20a and $10^{-2}$ for Eta Carinae.}
\label{fig8}
\end{figure*}

\subsection{Confrontation with observations of specific open clusters}

We calculate the $\gamma$-ray spectra produced by protons in hadronic collisions with the matter of the open cluster, applying their diffusion and interaction model as described above. As before, we consider the open clusters which contain the WR 20a and Eta Carinae binary systems. The calculations of the $\gamma$-ray spectra are performed for the same parameters of the acceleration scenario as discussed for the $\gamma$-ray production in the wind cavity regions of these two binaries. The results are shown for two different values of the parameter
$R_{20}^2B_{-4}n_{10}$ which determines the conditions within the open cluster. 

In the case of the open cluster containing the binary system WR 20a, the $\gamma$-ray spectrum expected from the open cluster Westerlund 2 is consistent with the positive detection of this open cluster by the HESS Collaboration (Abramowski et al.~2011), provided that the parameter $R_{20}^2B_{-4}n_{10} \le 1.5$ (see Fig.~8). Note that $\gamma$-rays produced in this case are expected to contribute mainly to the energy range around $\sim 1$ TeV. This is clearly below the energy range at which the $\gamma$-ray emission from the wind cavity is expected. $\gamma$-rays, expected from the open cluster in terms of our model for Westerlund 2, are predicted to be still detected by the CTA provided that the parameter $R_{20}^2B_{-4}n_{10} \ge 0.15$.

We have also compared the predictions of our model with the observations of the Carina Complex containing the Eta Carinae binary system. HESS Collaboration has derived an upper limit on the $\gamma$-ray flux from the extended source from this direction (Abramowski et al.~2012b), which is about a factor of five higher than the upper limit on the point source towards the Eta Carinae itself. We have compared the calculated $\gamma$-ray spectra with the HESS upper limit in Fig.~8. These spectra are still consistent with the upper limit provided that the value of the parameter $R_{20}^2B_{-4}n_{10} \le 0.2$. $\gamma$-rays produced in terms of our model should contribute only to the sub-TeV energy range. We conclude that the $\gamma$-ray emission, produced by relativistic protons in the Carina Complex, is expected to be still detectable by the CTA provided that the open cluster is characterized by the parameter $R_{20}^2B_{-4}n_{10} \ge 0.04$. However, this $\gamma$-ray component is expected to extend only through the low energy range of the CTA sensitivity, i.e. at $\sim$100 GeV.

\section{Neutrinos from the vicinity of binary system}

Detection of neutrinos, produced in hadronic interactions between relativistic protons and the matter of the stellar wind and/or open cluster, could provide additional constraints on the high energy processes in the considered scenario.
Therefore, we calculate the neutrino spectra from the wind cavity and the open cluster applying derived above spectra of relativistic protons in these regions.

The neutrino spectra from the wind cavity (on the left) and the open cluster (on the right) are shown in Fig.~9 for the WR 20a binary system (solid) and Eta Carinae binary system (dashed). They are obtained for the parameters of the model as used for the corresponding $\gamma$-ray spectra shown above, i.e. for the energy conversion inefficiencies as described in Fig.~6 and the distances to these binary systems equal to $\sim$ 2 kpc. These spectra are compared with the atmospheric neutrino background (ANB) and the upper limit on the neutrino flux from discrete sources reported by the ANTARES Collaboration (Adrian-Martinez et al.~2012) and the 5 yr sensitivity of the IceTop + IceCube (Aartsen et al.~2013). The neutrino spectra, expected from these two binary systems immersed within the open clusters, are on the level of the atmospheric neutrino background. However, they are about an order of magnitude below the present upper limit of the flux from discrete sources by ANTARES collaboration. It looks that nuclei are not energetic enough in order to produce neutrino fluxes which could be potentially detected by the Ice Cube telescope.
We have also calculated the expected neutrino flux produced by relativistic protons within the open cluster applying the normalization obtained from the comparison of the $\gamma$-ray fluxes with observations of these two open clusters (see Fig.~8).
These neutrino fluxes are predicted on much lower level than those expected from the wind cavities of these binary systems.
We conclude that considered binary systems are not expected to produce detectable neutrino fluxes from the directions of the open clusters Westerlund 2 and Carina Complex. Note however, that neutrino fluxes should be larger if the specific open cluster contains many massive binary systems. Therefore, our results do not completely exclude future detection of neutrinos from massive stellar clusters at distances of a few kpc from the vicinity of the Sun. 

\begin{figure*}
\vskip 6.5truecm
\includegraphics{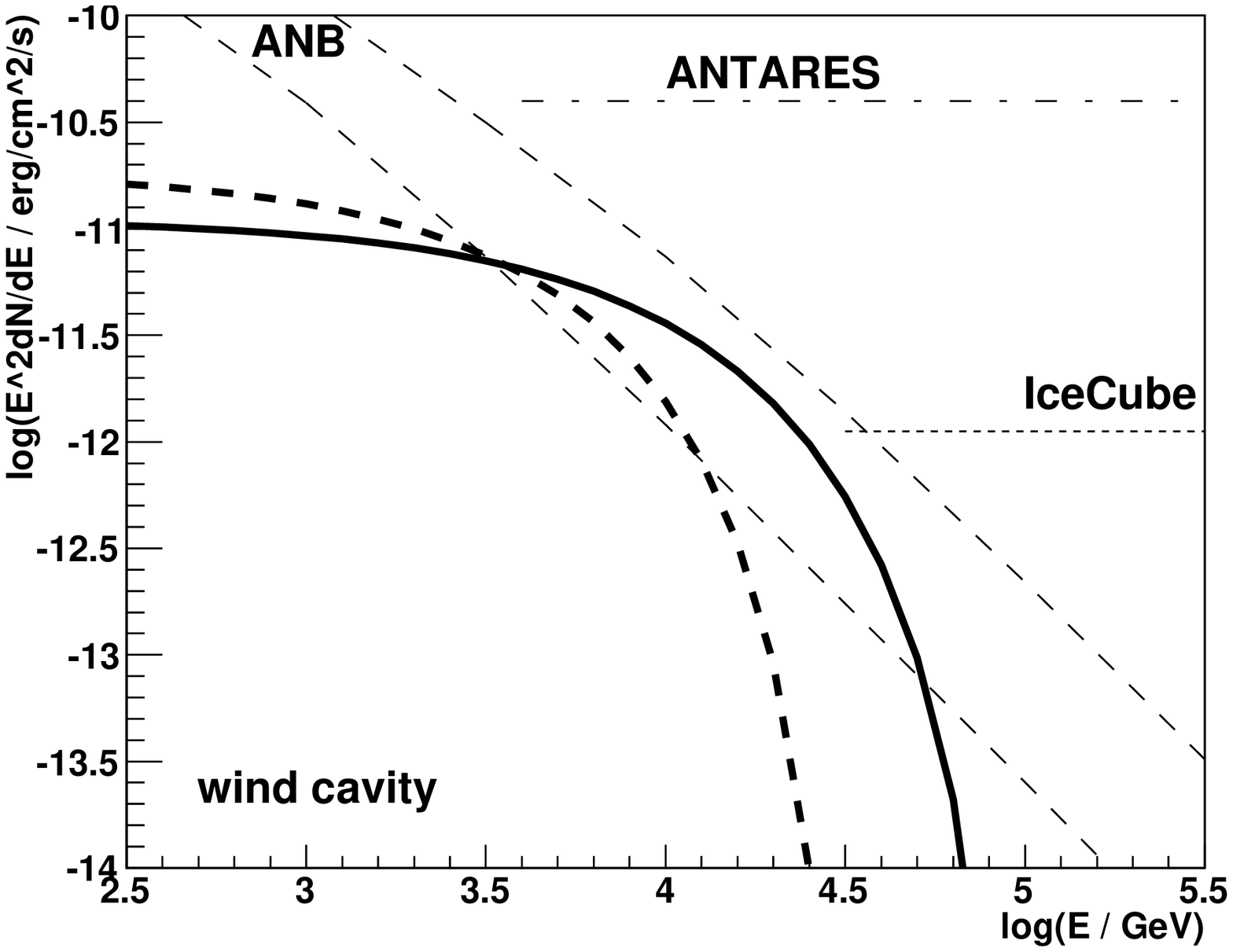}
\includegraphics{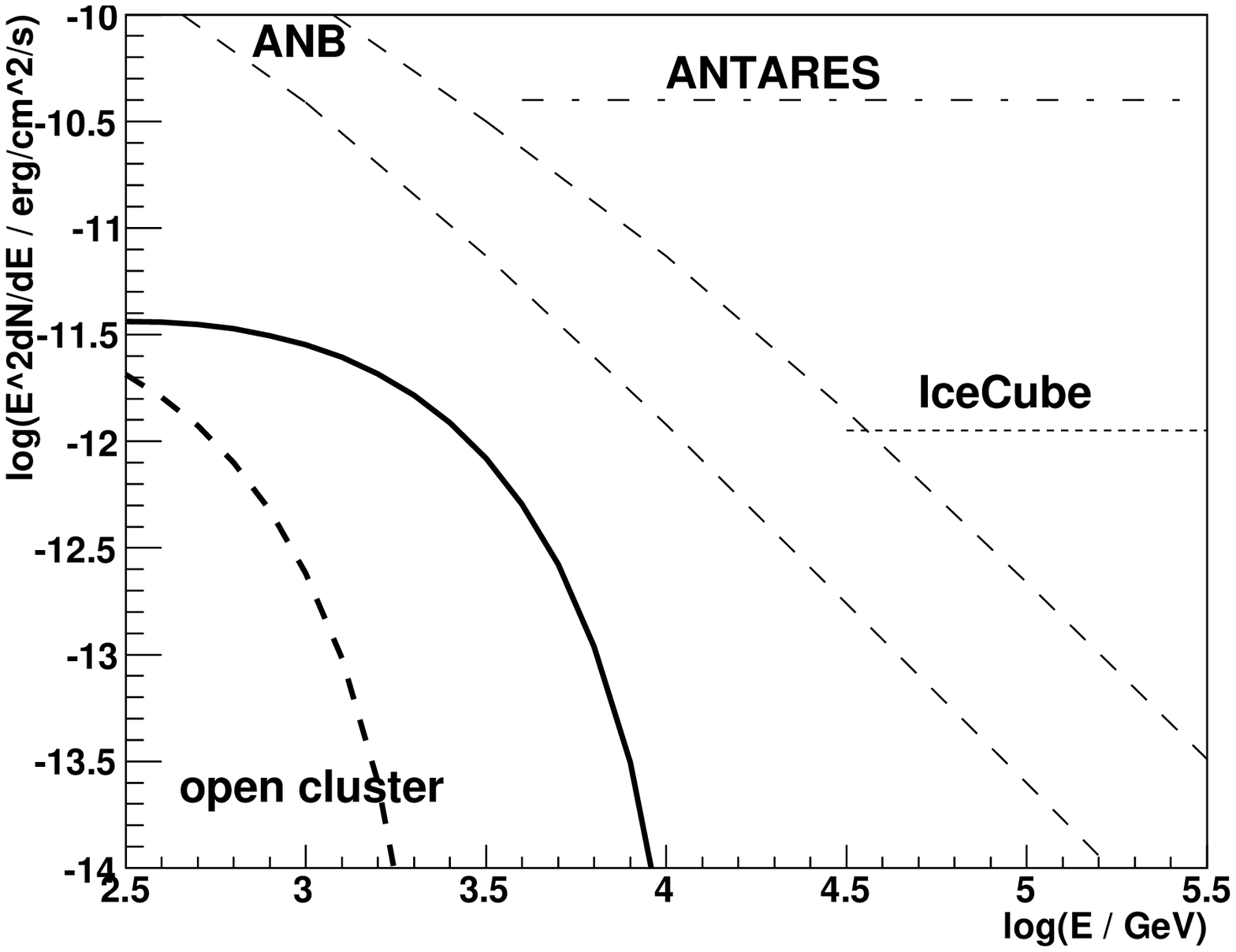}
\caption{Spectra of neutrinos (SED) produced by hadrons in the above discussed scenarios from the wind cavity (left) and from the open cluster (right). The minimum energy of escaping protons is equal to $10^5$ GeV, corresponding to $R_{20}^2B_{-4}n_{10} = 1.5$, for WR 20a (solid) and $1.5\times 10^4$ GeV, corresponding to $R_{20}^2B_{-4}n_{10} = 0.2$ for Eta Carinae (dashed). The neutrino spectra are produced by protons which are the decay products of neutrons extracted from nuclei in their collisions with stellar wind. The atmospheric background (ANB) in a viewcone of $1^\circ$ radius around the source is shown by the thin dashed curves (Abbasi et al.~2011), the 5 yr sensitivity of the IceTop + IceCube is shown by the thin dotted line (Aartsen et al.~2013), and the ANTARES upper limit on the point sources is shown by dot-dashed line (Adrian-Martinez et al.~2012). The proton spectra are normalized in this same way is described for the $\gamma$-ray spectra.}
\label{fig9}
\end{figure*}

\section{Conclusions}

Star forming regions contain plenty of extreme objects, such as massive binary systems and remnants of their evolution (supernova remnants, PWNe, X-ray binaries). These objects might be responsible for the acceleration of hadrons to TeV-PeV energies. In fact, GeV-TeV $\gamma$-ray emission has been already detected from a few open clusters. In this paper we concentrate on the processes due to the presence of massive binary systems in dense regions (open clusters). We have formulated a model for the interaction of the binary system with the matter of the open cluster. It is postulated that nuclei (from helium to oxygen) can be efficiently accelerated within the binary system. We show that these nuclei are disintegrated in the dense stellar radiation and the matter of the stellar wind injecting relativistic protons and neutrons. The fate of these secondary particles is followed in detail in the region of the stellar wind cavity and in the surrounding dense open cluster. We calculate the expected $\gamma$-ray and neutrino emission produced in the interaction of these particles with the matter of the stellar wind and surrounding open cluster.

The results of calculations are shown for the case of the two well known massive binary systems (WR 20a and Eta Carinae), which have been recently reported as a $\gamma$-ray sources in the GeV-TeV energy range. 
Our calculations show that the largest fluxes of $\gamma$-rays are produced by protons close to the binary system where the density of stellar wind is the largest. However, the absorption of gamma-rays in the stellar radiation field has the important effect on the gamma-ray spectrum, produced by protons extracted directly from the nuclei.
Therefore, the $\gamma$-ray spectra produced in these regions are clearly steeper than the injection spectra of protons (equal to the spectrum of accelerated nuclei). On the other hand, $\gamma$-ray spectra, from hadronic collisions of protons which are decay products of neutrons extracted form nuclei, are mainly produced at larger distances from the binary systems since neutrons move ballisticaly through the wind cavity and decay at relatively large distances. They interact with the matter of the wind with lower density but do not suffer strong absorption in the stellar radiation field.  In total, these $\gamma$-ray spectra have clearly lower level than spectra produced in previous process.

Significant amount of protons, which appeared in the wind cavity as a result of decay of neutrons, is advected with the stellar wind to the open cluster. We calculate the spectra of these protons taking into account their adiabatic and collisional energy losses. Protons, directly extracted from nuclei, cannot be advected to the open cluster with large energies due to the huge adiabatic energy losses. Therefore, the radiation produced by them can be safely neglected. Another population of relativistic protons is provided by the most energetic neutrons which can decay directly into the open cluster, i.e. outside the wind cavity. We take these protons into
account when calculating the radiation produced in the open cluster. The fate of relativistic protons in the open cluster depends on the cluster parameters. Protons diffuse outside the cluster and interact with the cluster matter. We show, that in the case of Bohm diffusion approximation, the escape/interaction conditions of protons depend on the parameter $R_{20}^2B_{-4}n_{10}$, which is the combination of the parameters characterizing the open cluster, its radius, magnetic field strength and density of matter. 
We show that for likely parameters of the open cluster, protons with largest expected energies escape from the cluster practically without interaction with the matter.
Only lower energy protons are captured in the open cluster and lose energy on the production of $\gamma$-rays and neutrinos. Note that protons escaping from the clusters are expected to have Lorentz factors in the range 
$\sim 10^{(4-5)}$. Therefore, we expect that open clusters, containing massive binary systems, might become interesting sources of relativistic protons in the Galaxy.

We confronted the $\gamma$-ray emission, expected in terms of this model, with the observations of the open clusters containing binary systems WR 20a (Westerlund 2) and Eta Carinae (Carina Complex). It is concluded that protons within the specific open cluster can contribute to the observed TeV $\gamma$-ray spectrum (mainly at its lower energy part) observed from Westerlund 2. This $\gamma$-ray emission is consistent with the upper limits on the TeV flux from Carina Complex. We also determine the minimum constraints on the parameters of these two clusters for which they will be detectable by the planned CTA.

We have also calculated the neutrino spectra expected in this model for these two binary systems. Unfortunately,
these neutrino fluxes are about two orders of magnitude below the present upper limit on neutrino emission from discrete sources provided by ANTARES Collaboration (Adrian-Martinez et al.~2012). These neutrinos will be also
difficult to observe with the IceCube telescope since their fluxes, produced in the wind cavity regions, are comparable to the atmospheric neutrino background. They are clearly below this background in the case of neutrinos produced in the open cluster itself.

In fact  many massive binary systems can be present at the same time in a specific open cluster. Therefore, in principle larger $\gamma$-ray and neutrino fluxes might be expected from specific open clusters
than calculated here fluxes from isolated binary systems. However the environment of the open cluster with many massive binary systems might be very complicated due to likely interaction between different wind cavities.
Therefore, simple re-scaling of the $\gamma$-ray and neutrino fluxes by the number of massive binaries in specific open cluster seems to be not adequate. A more complicated scenario should be considered in such a case.

\begin{acknowledgments}
We would like to thank the Referee for valuable comments.
This work is supported by the Polish Narodowe Centrum Nauki through the grant No. 2011/01/B/ST9/00411.
\end{acknowledgments}

\end{document}